\documentclass[aps,pra,amsmath,amssymb,amsfonts,reprint,twocolumn]{revtex4-2}
\usepackage[T1]{fontenc}
\usepackage{times}
\usepackage{siunitx}
\usepackage{graphicx}
\usepackage{dcolumn}
\usepackage{bm}
\usepackage{dcolumn}
\newcolumntype{d}[1]{D{.}{.}{#1}}

\newcommand*{\centt}[1]{\multicolumn{1}{c}{#1}}

\allowdisplaybreaks

\usepackage{xcolor}

\begin{document}
\preprint{Version 1.0}

\title{QED theory of the nuclear magnetic shielding  in H and $^3$He }

\author{Dominik Wehrli}
\email[]{dominik.wehrli@phys.chem.ethz.ch}
\affiliation{Laboratorium f\"ur Physikalische Chemie, ETH-Z\"urich, 8093 Z\"urich, Switzerland}

\author{Mariusz Puchalski}
\email[]{mpuchals@amu.edu.pl}
\affiliation{Faculty of Chemistry, Adam Mickiewicz University, Uniwersytetu Pozna{\'n}skiego 8, 61-614 Pozna{\'n}, Poland}

\author{Krzysztof Pachucki}
\email[]{krp@fuw.edu.pl} \homepage[]{www.fuw.edu.pl/~krp}

\affiliation{Faculty of Physics, University of Warsaw,
             Pasteura 5, 02-093 Warsaw, Poland}

\date{\today}

\begin{abstract}
The derivation of leading quantum electrodynamic corrections to the nuclear magnetic shielding in light hydrogen- and helium-like atomic systems
is described in detail. The presented theoretical approach applies to any light atomic and molecular systems, 
enabling the determination of the magnetic moment of light nuclei with much higher precision than known presently.
\end{abstract}
\maketitle

\section{Introduction}

The determination of the nuclear magnetic moment based on NMR spectra or atomic beam magnetic resonance measurements requires the calculation of the nuclear magnetic shielding constant \cite{jaszunski2012}. These calculations are usually performed using the Dirac-Coulomb Hamiltonian including the Breit interaction. 
If one aims for an accuracy as high as $\sim10^{-9}$, which is attainable experimentally \cite{WINELAND1983},
quantum electrodynamic (QED) effects should also be taken into account.
This accuracy, however, has not yet been reached in the calculation of the nuclear magnetic shielding, partly because of the difficulties with the calculation of QED effects.
There have been attempts \cite{koziol19} to include them within  the formalism based on the Dirac-Coulomb  Hamiltonian \cite{Kutzelnigg:2012},
but there is currently no adequate formulation of the QED theory for many electron systems.  Such a formulation exists only within the $1/Z$ expansion of the Hamiltonian \cite{SHABAEV2002}, 
where electron-electron interactions are treated perturbatively. 

For the one-electron systems (hydrogenic ions), Yerokhin {\em et al.} \cite{yerokhin11,yerokhin12} performed a nonperturbative numerical evaluation of one-loop QED contributions and observed a slow numerical convergence for the small nuclear charge $Z$. Therefore, these results were supplemented by the leading correction $\sim\alpha^5$ evaluated by nonrelativistic QED (NRQED) methods. However, some effects due to the magnetic moment anomaly were omitted there, resulting in small discrepancies compared to the nonperturbative results for the medium-$Z$ hydrogen-like ions. These discrepancies have been eliminated in our recent work \cite{wehrli:2021}. 

For helium, the leading QED logarithmic correction $\sim \alpha^5\,\ln\alpha$ was obtained by Rudzi\'nski {\em et al.} in Ref.~\cite{rudzinski09}, and the complete $\alpha^5$ correction was obtained in Ref. \cite{wehrli:2021}. 
Unexpectedly, significant cancellations were observed between the constant and the logarithmic  terms, resulting in a small QED correction for the magnetic shielding of about $96.3\cdot10^{-12}$.

In this work we present in detail the derivation of the leading QED corrections to the nuclear magnetic shielding in hydrogenic and helium-like systems. 
Most importantly, the obtained formulas can be generalized to any light few-electron system, which will enable the determination of
nuclear magnetic moments with significantly improved accuracy, for example $10^{-9}$ for $^9$Be from the measurement of the electron-nucleus magnetic moment ratio \cite{WINELAND1983}.

 \section{QED theory of the magnetic shielding}
The coupling of the nuclear magnetic moment $\vec\mu$ with the homogeneous magnetic field $\vec B$
is modified by the presence of atomic electrons according to \cite{ramsey50}
\begin{equation}
\delta H = -\vec\mu\cdot\vec B\,(1-\sigma). \label{eq:shielding_definition}
\end{equation}
We assume, what is particularly suited for light atomic systems, 
the expansion of the binding energy and the magnetic shielding $\sigma$ in the fine structure constant $\alpha$ and the electron-nucleus mass ratio $m/m_N$,

\begin{align}
\sigma =&\ \sigma^{(2)} + \sigma^{(4)} + \sigma^{(5)} + \sigma^{(6)} \label{02}
+\sigma^{(2,1)} + \sigma^{(2,2)} + \sigma^{(4,1)} + \ldots
\end{align}
The expansion terms $\sigma^{(n)}\propto\alpha^n$ are subsequently the nonrelativistic shielding,
the relativistic, the leading QED, and the higher-order QED corrections to the shielding in the infinite nuclear mass approximation. The terms 
$\sigma^{(n,k)}\propto\alpha^n\,(m/m_N)^k$ are the corrections due to the finite nuclear mass. 
In the case of one- \cite{ivanov09} and two-electron systems \cite{rudzinski09} all lower-order corrections are well-known, 
while the derivation of the leading QED correction is presented here, using the theory of nonrelativistic QED (NRQED) \cite{caswell86}.

Within the NRQED formalism, quantum electrodynamic  effects are incorporated in a general effective Hamiltonian \cite{zatorski10}. 
For the case of an electron subjected to electromagnetic fields $\vec{E}$ and $\vec{B}$, this effective Hamiltonian is given by
\begin{widetext}
\begin{align}
  H_{\text{NRQED}}={}&\frac{\vec{\pi}^{2}}{2\,m} -\frac{\vec{\pi}^4}{8\,m^3} +e\,A^0
  -\frac{e}{2\,m}\,(1+\kappa)\,\vec{\sigma}\cdot\vec{B} +\frac{e}{8\,m^3}\,\lbrace\vec{\pi}^2,\vec{\sigma}\cdot\vec{B}\rbrace
  -\frac{e^2}{2}\left(\frac{1}{4\,m^3}+\alpha_M\right)\vec{B}^2\nonumber\\
  {}& -\frac{e}{6}\left(\frac{3}{4\,m^2}+r^2_E + r^2_{\rm vp}\right)\vec\nabla\cdot\vec{E}
  -\frac{e}{8\,m^2}\,(1+2\,\kappa)\,\vec{\sigma}\cdot(\vec{E}\times\vec{\pi}-\vec{\pi}\times \vec{E})\nonumber\\
  {}& +\frac{e\,\kappa}{8\,m^3}\,\lbrace\vec{\pi}\cdot\vec{B},\vec{\sigma}\cdot\vec{\pi}\rbrace+\frac{e}{6\,m}
  \left(r^2_E+r^2_{\rm vp}-\frac{3\,\kappa}{4\,m^2}\right)(\vec\nabla\times \vec{B})\cdot\vec{\pi},
\label{eq:h_nrqed}
\end{align}
\end{widetext}
where we use $\hbar=c=\varepsilon_0=1$, $\{\,,\}$ denotes the anticommutator, $\vec \pi = \vec p - e\,\vec A$, $\kappa$ is the magnetic moment anomaly, and
\begin{align}
\kappa&=\frac{\alpha}{2\,\pi},\label{eq:kappa}\\
r_E^2&=\frac{3\,\kappa}{2\,m^2}+6\,F'_1(0) = 
\frac{2\,\alpha}{\pi\, m^2}\,\ln\frac{m}{\mu},\\
r^2_{\rm vp} &= -\frac{2\,\alpha}{5\,\pi\,m^2},\label{eq:rvp}\\
\alpha_M&=\frac{4\,\alpha}{3\,\pi\, m^3}\left(\frac{11}{8}-\ln\frac{m}{\mu}\right),
\end{align}
where $r_E$ and $r_{\rm vp}$ are the one-loop electron self-energy and the vacuum-polarization QED contributions to the charge radius, respectively, 
while $\alpha_M$ is the one-loop self-energy contribution to the magnetic polarizability \cite{yerokhin12}.
Without QED ($r^2_E=r^2_{\rm vp}=\alpha_M=\kappa=0$), the effective Hamiltonian
$H_{\rm NRQED}$ reduces to the nonrelativistic expansion of the 
Dirac Hamiltonian by the Foldy-Wouthuysen transformation \cite{itzykson05}.

When it comes to the QED parameters $r^2_E$ and $\alpha_M$,
it is convenient to use a photon momentum cut-off $\epsilon$ in the Coulomb gauge, rather than a finite photon mass $\mu$ as a regulator.
It can be shown that the two quantities are related to each other as \cite{itzykson05}
\begin{equation}
\ln\frac{\mu}{2\,\epsilon}=-\frac{5}{6}.
\end{equation}
With this substitution, these QED parameters read
\begin{align}
r^2_E &=\frac{2\,\alpha}{\pi\,m^2}\left(\ln\frac{m}{2\,\epsilon}+\frac{5}{6}\right),\label{eq:re2}\\
\alpha_M&=\frac{4\,\alpha}{3\,\pi\, m^3} \left(-\ln\frac{m}{2\,\epsilon}+\frac{13}{24}\right).\label{eq:alpham}
\end{align}
Their dependence on $\epsilon$ cancels out in the complete expression for any physical quantity, such as the Lamb shift
or the shielding constant, which will be demonstrated in the following. Although we are free to use any gauge,
as $H_{\rm NRQED}$ is gauge covariant, we will use the Coulomb gauge, as it is the most convenient gauge in studies of bound states.

The vector potential $\vec{A}$ in $H_\mathrm{NRQED}$ is the sum
of the external magnetic potential $\vec A_E$,
\begin{equation}
\vec A_E = \frac{1}{2}\,\vec B\times\vec r,
\end{equation}
and that due to the nuclear magnetic moment,
\begin{equation}  
\vec A_I = \frac{1}{4\,\pi}\,\vec\mu\times\frac{\vec r}{r^3}.
\end{equation}
Following Ramsey's theory of magnetic shielding \cite{ramsey50,helgaker99}, we split the Hamiltonian $H_{\text{NRQED}}$ as
\begin{align}
H_{\text{NRQED}}=&\ H_0 + \delta H,\\
\delta H=&\ \delta H_{\vec{A}_E=\vec{A}_I=0}+ \delta H_{\vec{A}_I,\vec{A}_E=0}+ \delta H_{\vec{A}_E,\vec{A}_I=0}
\nonumber \\ &\ 
+\delta H_{\vec{A}_E,\vec{A}_I}+O(\vec{A}^2_{I,E}),
\end{align}
where $\delta H$ is treated as a perturbation to the nonrelativistic Hamiltonian $H_0$,
$\delta H_{\vec{A}_E=\vec{A}_I=0}$ is independent of the magnetic fields,
$\delta H_{\vec{A}_I,\vec{A}_E=0}$ is linear in $\vec{A}_I$, $\delta H_{\vec{A}_E,\vec{A}_I=0}$
is linear in $\vec{A}_E$, and $\delta H_{\vec{A}_E,\vec{A}_I}$ is bilinear in both fields.
Because we are only interested in energy corrections that are proportional to $\vec{\mu}\cdot\vec{B}$, we write
\begin{align}
\delta E =&\ \langle\delta H\rangle + \left\langle\delta H\,\frac{1}{(E_0-H_0)'}\,\delta H\right\rangle\nonumber\\
=&\ \langle\delta H_{\vec{A}_E,\vec{A}_I}\rangle + 2\left\langle\delta H_{\vec{A}_E,\vec{A}_I}\,\frac{1}{(E_0-H_0)'}\,\delta H_{\vec{A}_E=\vec{A}_I=0}\right\rangle\nonumber \\ &
+ 2\left\langle\delta H_{\vec{A}_I,\vec{A}_E=0}\,\frac{1}{(E_0-H_0)'}\,\delta H_{\vec{A}_E,\vec{A}_I=0}\right\rangle+\ldots,
\label{eq:ecorrectiongeneral}
\end{align}
where 
\begin{equation}
\frac{1}{(E_0-H_0)'}\equiv\frac{1}{E_0-H_0}(1-|\phi_0\rangle\langle\phi_0|)
\end{equation}
is the reduced Green's function, and the ellipsis denotes terms that are not proportional to $\vec{\mu}\cdot\vec{B}$ and will be discarded. 
All the expectation values are taken with respect to the ground state $\phi_0$ of $H_0$.
The spherical symmetry of $\phi_0$ then implies the relation
\begin{equation}
\mu^iB^j=\frac{\delta^{ij}}{3}\,\vec{\mu}\cdot\vec{B},\label{eq:mubsymmetry}
\end{equation}
which allows for a simple factoring of $\vec{\mu}\cdot\vec{B}$ from many terms appearing in Eq.\,\eqref{eq:ecorrectiongeneral},
and the magnetic shielding constant $\sigma$ is obtained
through the relation $\delta E = \sigma\,\vec{\mu}\cdot\vec{B}$.

\section{Magnetic shielding in hydrogen-like ions with small nuclear charge}
\subsection{Leading-order $\alpha^2$ and relativistic $\alpha^4$ contributions}

Let us assume an electron placed in the field of an infinitely heavy point nucleus, 
i.e., $A^0 = -Z\,e/(4\,\pi\,r)$ with  $\vec E=-\vec \nabla A^0$.
The first derivation of the leading-order magnetic shielding for atoms and molecules was presented
by Ramsey \cite{ramsey50}. Later, the Dirac equation was used to calculate the shielding constant
for hydrogen-like ions to incorporate relativistic effects to all orders \cite{ivanov09}.
In this section we rederive the nonrelativistic and the leading relativistic correction to the magnetic shielding of hydrogen-like ions, 
which, in contrast to the Dirac formalism, allows a straightforward generalization to many-electron systems.

The nonrelativistic shielding constant comes from the $\vec A^2$ term in the electron kinetic energy $\vec{\pi}^2/(2\,m)$
in Eq.\,\eqref{eq:h_nrqed}.
Thus, the relevant energy correction is
\begin{align}
E^{(2)}
\approx &\ \frac{e^2}{2\,m}\,\langle \vec A^2\rangle
\approx \frac{e^2}{m}\,\langle \vec A_E \cdot \vec A_I\rangle \nonumber \\
=&\ \frac{\alpha}{2\,m}\,\biggl\langle (\vec B\times\vec r)
\cdot \left(\vec\mu\times\frac{\vec r}{r^3}\right)\biggr\rangle,
\end{align}
where the matrix elements are calculated with the nonrelativistic wave function, being the ground state
of the nonrelativistic Hamiltonian $H_0$,
\begin{align}
  H_0 =\frac{\vec{p}^{\,2}}{2\,m} -\frac{Z\,\alpha}{r}.\label{eq:h0}
\end{align}
Using Eq.\,\eqref{eq:mubsymmetry} we obtain the shielding constant for the hydrogenic ground state as
\begin{equation}
\sigma^{(2)} = \frac{\alpha}{3\,m}\,\biggl\langle
\frac{1}{r}\biggr\rangle = \frac{1}{3}\,\alpha\,(Z\,\alpha).
\label{eq:sigma2hydrogen}
\end{equation}

For the derivation of the relativistic correction we note that terms that are proportional to the angular momentum $\vec{L}$ 
vanish because we only consider corrections to the ground state. Keeping only terms of order $\alpha^4$, Eq.\,\eqref{eq:ecorrectiongeneral} becomes
\begin{widetext}
\begin{align}
E^{(4)}={}&  \left\langle-\frac{e^2}{4\,m^3}\,\lbrace\vec{p}^{\,2},\vec{A}_E\cdot\vec{A}_I\rbrace- \frac{e^2}{4\,m^3}\,\vec{B}\cdot\vec{B}_I\right\rangle
+2\left\langle\frac{e^2}{m}\,\vec{A}_E\cdot\vec{A}_I\, \frac{1}{(E_0-H_0)'} \left(\frac{\pi\, Z\,\alpha}{2\,m^2}\, \delta(\vec{r})-\frac{\vec{p}^{\,4}}{8\,m^3}\right)\right\rangle\nonumber\\
{}& + 2\left\langle\left[\frac{e}{8\,m^3}\,\lbrace \vec{p}^{\,2},\vec{\sigma}\cdot\vec{B}\rbrace - \frac{e\, Z\,\alpha}{4\,m^2}\,\vec{\sigma}\cdot\left(\frac{\vec{r}}{r^3}\times\vec{A}_E\right)\right] \frac{1}{(E_0-H_0)'}\left(-\frac{e}{2\,m}\vec{\sigma}\cdot\vec{B}_I\right)\right\rangle,
\label{eq:e4start}
\end{align}
where $\vec{B}_I=\nabla\times\vec{A}_I$.
This can be simplified and written in terms of the shielding constant as
\begin{align}
\sigma^{(4)}={}&-\frac{\alpha}{6\,m^3}\left\langle\frac{1}{r}\,\vec{p}^{\,2}\right\rangle-\frac{2\,\pi\,\alpha}{3\,m^3}\,\langle\delta(\vec{r})\rangle + \frac{5\,\pi\,\alpha\,(Z\,\alpha)}{9\,m^3}\left\langle\frac{1}{r}\,\frac{1}{(E_0-H_0)'}\,\delta(\vec{r})\right\rangle-\frac{\alpha}{12\,m^4}\left\langle\frac{1}{r}\,\frac{1}{(E_0-H_0)'}\,\vec{p}^{\,4}\right\rangle\nonumber\\
{}&-\frac{2\,\pi\,\alpha}{3\,m^3}\left\langle\delta(\vec{r})\,\frac{1}{(E_0-H_0)'}\,\vec{p}^{\,2}\right\rangle -\frac{\alpha\,(Z\,\alpha)}{8\,m^3}\left\langle\left(\frac{r^i\,r^j}{r^5}\right)^{(2)}\frac{1}{E_0-H_0}\left(\frac{r^i\,r^j}{r^3}\right)^{(2)}\right\rangle,
\end{align}
\end{widetext}
where  the second rank tensors are defined by  
\begin{equation}
(p^iq^j)^{(2)}=\frac{p^iq^j}{2}+\frac{p^jq^i}{2}-\vec{p}\cdot\vec{q}\,\frac{\delta^{ij}}{3}.
\end{equation}
Using the hydrogenic matrix elements from Table \ref{tab:h_expvalues} we obtain
\begin{align}
\sigma^{(4)}= \frac{97}{108}\,\alpha\,(Z\,\alpha)^3\,. \label{eq:sigma4hydrogen}
\end{align}
This result my be compared with the correction found from the Dirac equation \cite{ivanov09},
\begin{align}
\sigma &= -\frac{4}{9}\alpha\, (Z\,\alpha)\left[\frac{1}{3}-\frac{1}{6\,(1+\gamma)}+\frac{2}{\gamma}-\frac{3}{2\,\gamma-1}\right]\nonumber\\
&=\alpha\, (Z\,\alpha)\left[\frac{1}{3}+\frac{97}{108}(Z\,\alpha)^2+\frac{289}{216}(Z\,\alpha)^4+\ldots\right],
\label{eq:diracexpansion}
\end{align}
where
\begin{equation}
\gamma=\sqrt{1-(Z\,\alpha)^2}.
\end{equation}
In Eq.\,\eqref{eq:diracexpansion} we recognize the result for $\sigma^{(2)}$ in Eq.\,\eqref{eq:sigma2hydrogen} and for $\sigma^{(4)}$ in Eq.\,\eqref{eq:sigma4hydrogen}
as the first two terms of the expansion, as expected.

\begin{table}[ht]
	\caption{Formulas for ground state expectation values in hydrogenic systems.}\label{tab:h_expvalues}
	\begin{center}
		\begin{tabular}{ll}
			\hline \hline \\
			Operator &  Expectation value \\[0.2em]
			\hline\\
			$\delta(\vec{r})$ & $\dfrac{1}{\pi}\,(Z\,\alpha)^3\,m^3$\\[1em]
			$\vec{p}^{\,2}$ & $(Z\,\alpha)^2\,m^2$\\[1em]
			$\dfrac{1}{r}\,\vec{p}^{\,2}$ & $3\,(Z\,\alpha)^3\,m^3$\\[1em]
			$\dfrac{1}{r}\,\dfrac{1}{(E_0-H_0)'}\,\delta(\vec{r})$ & $-\dfrac{3}{2\,\pi}\,(Z\,\alpha)^2\,m^3$\\[1em]
			$\dfrac{1}{r}\,\dfrac{1}{(E_0-H_0)'}\,\vec{p}^{\,2}$ & $-Z\,\alpha\,m^2$\\[1em]
			$\delta(\vec{r})\,\dfrac{1}{(E_0-H_0)'}\,\vec{p}^{\,2}$ & $-\dfrac{3}{\pi}\,(Z\,\alpha)^3\,m^4$\\[1em]
			$\dfrac{1}{r}\,\dfrac{1}{(E_0-H_0)'}\,\vec{p}^{\,4}$ & $-10\,(Z\,\alpha)^3\,m^4$\\[1em]
			$\delta(\vec{r})\,\dfrac{1}{(E_0-H_0)'}\,\vec{p}^{\,2}$ & $-\dfrac{3}{\pi}\,(Z\,\alpha)^3\,m^4$\\[1em]
			$\left(\dfrac{r^i\,r^j}{r^5}\right)^{(2)}\dfrac{1}{E_0-H_0}\left(\dfrac{r^i\,r^j}{r^2}\right)^{(2)} $ & $-\dfrac{8}{27}\,(Z\,\alpha)\,m^2$\\[1em]
			$\left(\dfrac{r^i\,r^j}{r^5}\right)^{(2)}\dfrac{1}{E_0-H_0}\left(\dfrac{r^i\,r^j}{r^3}\right)^{(2)} $ & $-\dfrac{14}{27}\,(Z\,\alpha)^2\,m^3$\\[1em]
			\hline\hline
		\end{tabular}
	\end{center}
\end{table}

\subsection{ $\alpha^5$ QED correction without magnetic field}
Before considering QED effects to the shielding constant, let us first recalculate them for hydrogenic energy levels 
with vanishing angular momentum, following Ref. \cite{pachucki98}.
The leading QED correction (Lamb shift) to hydrogenic energy levels is obtained by splitting it into the low- and high-energy parts,
\begin{equation}
E^{(5)} =  E_L^{(5)} + E_H^{(5)}.
\end{equation}
The high-energy part is the following expectation value of the relevant terms from the NRQED Hamiltonian in Eq.\,\eqref{eq:h_nrqed},
\begin{align}
E_H^{(5)} =&\ \biggl\langle
-\frac{e}{6}\,(r_E^2 + r^2_{\rm vp})\,\vec\nabla\cdot\vec E\biggr\rangle
\nonumber \\ =&\
\frac{2}{3}\,\frac{m\,(Z\,\alpha)^4}{n^3}\,(r_E^2 + r^2_{\rm vp})
\nonumber \\=&\ 
\frac{\alpha}{\pi}\,\frac{m\,(Z\,\alpha)^4}{n^3}\,
\biggl(\frac{4}{3}\,\ln\frac{m}{2\,\epsilon}+\frac{10}{9}-\frac{4}{15}\biggr)\,,
\label{eq:ehhydrogenlambshift}
\end{align}
while the low-energy part (in the Coulomb gauge) is due to emission and absorption
of the low-energy $(k<\epsilon)$ photons,
\begin{align}
E_L^{(5)} =&\ e^2\int_{k<\epsilon}
\frac{d^3k}{(2\,\pi)^3\,2\,k}\,
\biggl(\delta^{ij}-\frac{k^i\,k^j}{k^2}\biggr)\nonumber \\ &\ \times
\biggl\langle\frac{p^i}{m}\,
\frac{1}{E_0-H_0-k}
\,\frac{p^j}{m}\biggr\rangle\nonumber \\ = &\
\frac{2\,\alpha}{3\,\pi}\,
\biggl\langle\frac{\vec p}{m}\,(H_0-E_0)\,\biggl\{
\ln\biggl[\frac{2\,\epsilon}{m\,(Z\,\alpha)^2}\biggr]
\nonumber \\ &\ 
-\ln\biggl[\frac{2\,(H_0-E_0)}{m\,(Z\,\alpha)^2}\biggr]\biggr\}
\frac{\vec p}{m}\biggr\rangle.
\label{eq:ellamb}
\end{align}
The total Lamb shift thus becomes
\begin{align}
E^{(5)} =&\ \frac{\alpha}{\pi}\,\frac{m\,(Z\,\alpha)^4}{n^3}\,
\biggl\{\frac{4}{3}\ln\bigl[(Z\,\alpha)^{-2}\bigr]
+\biggl(\frac{10}{9}-\frac{4}{15}\biggr)
\nonumber \\ &\ 
-\frac{4}{3}\,\ln k_0(n)\biggr\},
\label{eq:hydrogenlambshift}
\end{align} 
where
\begin{align}
\ln k_0(n) = \frac{n^3}{2\,m^3\,(Z\,\alpha)^4}\,
\biggl\langle\!\vec p\,(H_0-E_0) \ln\biggl[\frac{2\,(H_0-E_0)}{m\,(Z\,\alpha)^2}\biggr] \vec p \biggr\rangle
\label{eq:betheloghydrogen}
\end{align}
is the Bethe logarithm \cite{itzykson05,bethe77,bethe47}. The $\ln\epsilon$ term cancels out, as expected.

\subsection{$\alpha^5$ QED correction to the magnetic shielding}\label{sec:hydroqedcorr}
The derivation of the QED correction to the magnetic shielding in hydrogenic systems
using the formalism of NRQED was first presented in Refs.\ \cite{yerokhin11,yerokhin12}, 
omitting accidentally some contributions due to the electron magnetic moment anomaly.
Here we present a derivation of the complete QED correction, which is in most parts similar to that in Ref. \cite{yerokhin12}.
In analogy to the derivation of the Lamb shift, we split the correction into low- and high-energy contributions,
\begin{equation}
E^{(5)}=E^{(5)}_L+E^{(5)}_H,
\end{equation}
where $E^{(5)}_H$ is given by
\begin{widetext}
\begin{align}
E^{(5)}_H={}& \left\langle-e^2\,\alpha_M\,\vec{B}\cdot\vec{B}_I\right\rangle +\left\langle\frac{e}{6\,m}\left(r^2_E+r^2_{\rm vp}-\frac{3\,\kappa}{4\,m^2}\right)(\nabla\times \vec{B}_I)\cdot(-e\,\vec{A}_E)\right\rangle\nonumber\\
{}&+2\,\left\langle\frac{e^2}{m}\,\vec{A}_E\cdot\vec{A}_I\, \frac{1}{(E_0-H_0)'} \left[\frac{2\,\pi\, Z\,\alpha}{3}\,(r^2_E+r^2_{\rm vp})\,\delta(\vec{r})\right]\right\rangle\nonumber\\ 
{}&+2\,\left\langle\left[\frac{e\,\kappa}{8\,m^3}\,\lbrace \vec{p}^{\,2},\vec{\sigma}\cdot\vec{B}\rbrace 
-\frac{3\,e\,Z\alpha}{4\,m^2}\,\kappa\,\vec{\sigma}\cdot\left(\frac{\vec{r}}{r^3}\times\vec{A}_E\right)
+\frac{e\,\kappa}{8\,m^3}\,\lbrace\vec{p}\cdot\vec{B},\vec{\sigma}\cdot\vec{p}\rbrace\right] \frac{1}{(E_0-H_0)'}\left(-\frac{e}{2\,m}\vec{\sigma}\cdot\vec{B}_I\right)\right\rangle.\label{eq:e5hhydrogen}
\end{align}
The second line in Eq.\,\eqref{eq:e5hhydrogen} corresponds to the high-energy contribution of the Lamb shift in Eq.\,\eqref{eq:ehhydrogenlambshift}, 
and here in addition it includes wave function corrections due to the magnetic fields. Equation\,\eqref{eq:e5hhydrogen} leads to the following result for the high-energy part of the shielding,
\begin{align}
  \sigma_{H}^{(5)}={}&\frac{4\,\alpha^2}{3\,m^3}\left[\frac{5}{3}\ln\frac{m}{2\,\epsilon}
    -\left(\frac{301}{144}-\frac{1}{5}\right)\right]\langle\delta(\vec{r})\rangle
  + \frac{8\,\alpha^2\,(Z\,\alpha)}{9\,m^3}\left[\ln\frac{m}{2\,\epsilon}
  +\left(\frac{29}{24}-\frac{1}{5}\right)\right]\left\langle\frac{1}{r}\,\frac{1}{(E_0-H_0)'}\,\delta(\vec{r})\right\rangle\nonumber\\
  {}&
  -\frac{4\,\alpha^2}{9\,m^4}\left\langle\vec{p}^{\,2}\,\frac{1}{(E_0-H_0)'}\,\delta(\vec{r})\right\rangle
  +\frac{\alpha^2}{8\,\pi\, m^2}\left\langle\left(\frac{r^i\,r^j}{r^5}\right)^{(2)}\frac{1}{(E_0-H_0)'}
   \left[ -p^i\,p^j-\frac{3\,Z\,\alpha}{2\,m}\,\frac{r^i\,r^j}{r^3}\right]\right\rangle.
\end{align}
\end{widetext}
Using the hydrogenic matrix elements from Table \ref{tab:h_expvalues} we obtain
\begin{align}
  \sigma_{H}^{(5)} = \frac{8}{9\,\pi}\,\alpha^2\,(Z\,\alpha)^3\left[\ln\frac{m}{2\,\epsilon} - \frac{221}{64}+\frac{3}{5}\right].
\end{align}
The term $3/5$ in the brackets incorporates the vacuum polarization.

For the calculation of the low-energy part $E^{(5)}_L$,  we first define the nonrelativistic Hamiltonian in the  presence of the magnetic field
\begin{equation}
H=\frac{\vec{\pi}^2}{2\,m}-\frac{Z\,\alpha}{r},
\label{eq:hbfield}
\end{equation}
with the ground state energy $E$ that accounts for interaction with the magnetic fields. The low-energy contribution then reads
\begin{align}
E^{(5)}_L={}&e^2\int_{k<\epsilon}\frac{d^3k}{2\,k\,(2\,\pi)^3}\left(\delta^{ij}-\frac{k^i\,k^j}{k^2}\right)\\
{}&\times\left\langle\frac{\pi^i}{m}\,\frac{1}{E-H-k}\,\frac{\pi^j}{m}\right\rangle_{\!B}\nonumber\\
={}&\frac{2\,\alpha}{3\,\pi}\int_{0}^{\epsilon}dk\,k\left\langle\frac{\vec{\pi}}{m}\,\frac{1}{E-H-k}\,\frac{\vec{\pi}}{m}\right\rangle_{\!B},
\label{eq:e5lstart}
\end{align}
where the subscript $B$ indicates the expectation value over the ground state of $H$ in Eq.\,\eqref{eq:hbfield}. Evaluation of the integral in Eq.\,\eqref{eq:e5lstart}, and writing the terms with and without $\ln\epsilon$ separately, yields
\begin{align}
E^{(5)}_L&=E^{(5)}_{LA}+E^{(5)}_{LB},\label{eq:e5split}\\
E^{(5)}_{LA} &= -\frac{2\,\alpha}{3\,\pi}\left\langle\frac{\vec{\pi}}{m}\,(H-E)\ln\frac{2\,(H-E)}{m\,(Z\,\alpha)^2}\,\frac{\vec{\pi}}{m}\right\rangle_{\!B},\\
E^{(5)}_{LB} &=\frac{2\,\alpha}{3\,\pi}\left\langle\frac{\vec{\pi}}{m}\,(H-E)\,\frac{\vec{\pi}}{m}\right\rangle_{\!B}\ln\frac{2\,\epsilon}{m\,(Z\,\alpha)^2}.
\end{align}
We first turn to the calculation of $E^{(5)}_{LB}$ using the following identity,
\begin{equation}
\begin{split}
{}&2\left\langle\!\frac{\vec{\pi}}{m}(H-E)\frac{\vec{\pi}}{m}\right\rangle_{\!B} = \left\langle\left[\frac{\vec{\pi}}{m},\left[(H-E),\frac{\vec{\pi}}{m}\right]\right]\right\rangle_{\!B}\\
  ={}& \biggl\langle \frac{4\,\pi\, Z\,\alpha}{m^2}\,\delta(\vec{r})+\frac{e}{m^3}\nabla\!\times\!(\vec{B}+\vec{B}_I)\cdot\vec{\pi}+\frac{2\,e^2}{m^3}\,(\vec{B}+\vec{B}_I)^2\!\biggr\rangle_{\!B}
  \label{eq:doublecommtrick}
\end{split}
\end{equation}
and approximate the ground state $|\phi\rangle$ of $H$ by
\begin{equation}
|\phi\rangle=|\phi_0\rangle+\frac{1}{(E_0-H_0)'}\,\frac{e^2}{m}\,\vec{A}_E\cdot\vec{A}_I|\phi_0\rangle,\label{eq:phiapprox}
\end{equation}
where $|\phi_0\rangle$ is the ground state of the Hamiltonian $H_0$ without magnetic fields, 
as defined in Eq.\,\eqref{eq:h0}. Remembering that we only need to keep terms proportional to $\vec{\mu}\cdot\vec{B}$, 
we find with Eqs.\,\eqref{eq:doublecommtrick}~and~\eqref{eq:phiapprox} 
\begin{align}
  E^{(5)}_{LB}=&\ \frac{\alpha}{3\,\pi}\ln\frac{2\,\epsilon}{m\,(Z\,\alpha)^2}\,\biggl[\frac{8\,\pi\, Z\,\alpha^2}{3\,m^3}\,\vec{\mu}\cdot\vec{B}\left\langle\delta(\vec{r})\,\frac{1}{(E_0-H_0)'}\,\frac{1}{r}\right\rangle\nonumber \\ &\
-\frac{4\,\pi\,\alpha}{m^3}\,\langle\vec{A}_E\cdot(\nabla\times\vec{B}_I)\rangle+\frac{16\,\pi\,\alpha}{m^3}\,\langle\vec{B}\cdot\vec{B}_I\rangle\biggr]\nonumber\\
=&\ \frac{8}{9\,\pi}\ln\frac{2\,\epsilon}{m\,(Z\,\alpha)^2}\,\alpha^2\,(Z\,\alpha)^3\,\vec{\mu}\cdot\vec{B},
\label{eq:e5lb}
\end{align}
where all expectation values in the above are calculated with respect to $|\phi_0\rangle$.

For the derivation of $E^{(5)}_{LA}$ we start again from Eq.\,\eqref{eq:e5lstart}, transform it now in a different way, 
and finally drop all the terms including $\ln \epsilon$ because they are already included in $E^{(5)}_{LB}$. Noting that
\begin{equation}
\vec{\pi}=-i\,m\,[\vec{r},H-E],
\end{equation}
we write Eq.\,\eqref{eq:e5lstart} as
\begin{align}
E^{(5)}_{L}&=\frac{2\,\alpha}{3\,\pi}\int_{0}^{\epsilon}dk\,k\left\langle\frac{\vec{\pi}}{m}\,\frac{1}{E-H-k}\,\frac{\vec{\pi}}{m}\right\rangle_{\!B}\nonumber\\
&=\frac{2\,\alpha}{3\,\pi}\int_{0}^{\epsilon}dk\,k^3\left\langle\vec{r}\,\frac{1}{E-H-k}\,\vec{r}\right\rangle_{\!B}.
\label{eq:e5linterm}
\end{align}
The integrand in Eq.\,\eqref{eq:e5linterm} may be expanded in the magnetic fields by writing
\begin{equation}
H=H_0+\frac{\alpha}{3\,m\,r}\,\vec{\mu}\cdot\vec{B}-\frac{e}{2\,m}\,\vec{L}\cdot\vec{B}-\frac{e}{4\,\pi\, m\,r^3}\,\vec{L}\cdot\vec{\mu}+\ldots,
\end{equation}
where the ellipsis denotes terms that are not relevant for the shielding. 
The first term including the external field $\vec{B}$ can be absorbed into the nuclear charge via $Z'=Z-\vec{\mu}\cdot\vec{B}/(3\,m)$.
Defining $H_0'=\vec{p}^{\,2}/(2\,m)-Z'\,\alpha/r$ with the ground-state energy $E_0'$,  we rewrite Eq.\,\eqref{eq:e5linterm} as
\begin{align}
E^{(5)}_{L} = &\ E^{(5)}_{A1}+E^{(5)}_{A2},\\
E^{(5)}_{A1} =&\ \frac{2\,\alpha}{3\,\pi}\int_{0}^{\epsilon}dk\,k^3\left\langle\vec{r}\,\frac{1}{E_0'-H_0' -k}\,\vec{r}\right\rangle\,,\\
E^{(5)}_{A2} ={}& \frac{2\,\alpha^2}{3\,\pi\,m^2}\int_{0}^{\epsilon}dk\,k^3\,\biggl\langle\vec{r}\,\frac{1}{E_0-H_0 -k}\,\vec{L}\cdot\vec{B}\nonumber \\ &\ \times
\frac{1}{E_0-H_0 -k}\, \frac{\vec{L}\cdot\vec{\mu}}{r^3}\,\frac{1}{E_0-H_0 -k}\,\vec{r}\biggr\rangle\biggr].
\end{align}
We note that perturbations containing $\vec{L}$ do not change the ground-state energy or the ground-state wave function;
therefore, in $E^{(5)}_{A2}$ we now have the ground-state energy $E_0$ of $H_0$ rather than $E$, 
and the expectation value is evaluated for the ground state $|\phi_0\rangle$ of $H_0$. For the  $E^{(5)}_{A1}$ term we get
\begin{align}
E^{(5)}_{A1}&= \frac{2\,\alpha}{3\,\pi}\int_{0}^{\epsilon}dk\,k\left\langle\frac{\vec{p}}{m}\,\frac{1}{E'_0-H'_0-k}\,\frac{\vec{p}}{m}\right\rangle\nonumber\\
&= \frac{4}{3\,\pi}\,\alpha\,(Z'\,\alpha)^4\,m\left[\ln\frac{2\,\epsilon}{m\,(Z'\,\alpha)^2}-\ln k_0\right]\nonumber\\
&=\frac{16}{9\,\pi}\,\alpha^2\,(Z\,\alpha)^3\,\vec{\mu}\cdot\vec{B}\left[\ln k_0 +\frac{1}{2}-\ln\frac{2\,\epsilon}{m\,(Z\,\alpha)^2}\right],
\label{eq:e5l1}
\end{align}
where $\ln k_0\equiv\ln k_0(1)$ is the Bethe logarithm given in Eq.\,\eqref{eq:betheloghydrogen}.
Using $[\vec{L},H_0]=0$ and $[\vec{L}^2,r^i]|\phi_0\rangle=2\,r^i|\phi_0\rangle$, $E^{(5)}_{A2}$ becomes
\begin{widetext}
\begin{align}
E^{(5)}_{A2}&=\frac{4\,\alpha^2}{9\,\pi\, m^2}\,\vec{\mu}\cdot\vec{B}\int_{0}^{\epsilon}dk\,k^3 \left\langle\vec{r}\,\frac{1}{(E_0-H_0 -k)^2}\,\frac{1}{r^3}\,\frac{1}{E_0-H_0 -k}\,\vec{r}\right\rangle\nonumber\\
&=\frac{2\,\alpha^2}{9\,\pi\, m^2}\,\vec{\mu}\cdot\vec{B}\int_{0}^{\epsilon}dk\,k^3\, \frac{d}{dk}\left\langle\vec{r}\,\frac{1}{E'_0-H'_0 -k}\,\frac{1}{r^3}\,\frac{1}{E_0-H_0 -k}\,\vec{r}\right\rangle\nonumber\\
&=\frac{2\,\alpha^2}{9\,\pi\, m^2}\,\vec{\mu}\cdot\vec{B} \left[\epsilon^3\left\langle\vec{r}\,\frac{1}{E_0-H_0 -\epsilon}\,\frac{1}{r^3}\,\frac{1}{E_0-H_0 -\epsilon}\,\vec{r}\right\rangle - 3\int_{0}^{\epsilon}dk\,k^2\left\langle\vec{r}\,\frac{1}{E_0-H_0 -k}\,\frac{1}{r^3}\,\frac{1}{E_0-H_0 -k}\,\vec{r}\right\rangle \right].
\label{eq:el2partialint}
\end{align}
We expand the integrand in the first term initially in $\alpha$,
\begin{equation}
\frac{1}{E_0-H_0 -\epsilon} = - \frac{1}{\epsilon}-\frac{E_0-H_0}{\epsilon^2},
\label{eq:epsexpansione0h0}
\end{equation}
and subsequently take the limit $\epsilon\rightarrow0$, so that
\begin{align}
\epsilon^3\left\langle\vec{r}\,\frac{1}{E_0-H_0 -\epsilon}\,\frac{1}{r^3}\,\frac{1}{E_0-H_0 -\epsilon}\,\vec{r}\right\rangle
&=\left\langle\left[\left[\vec{r},(E_0-H_0)\right],\frac{\vec{r}}{r^3}\right]\right\rangle =-4\,(Z\,\alpha)^3\,m^2.
\end{align}
With the implicit definition
\begin{equation}
\int_{0}^{\epsilon}dk\,k^2\left\langle\vec{r}\,\frac{1}{E_0-H_0 -k}\,\frac{1}{r^3}\,\frac{1}{E_0-H_0 -k}\,\vec{r}\right\rangle = \epsilon\left\langle\frac{1}{r}\right\rangle - 4\,(Z\,\alpha)^3\,m^2\left[ \ln\frac{2\,\epsilon}{m\,(Z\,\alpha)^2}- \ln k_3\right]
\end{equation}
\end{widetext}
of the Bethe-type logarithm $\ln k_3$, we find
\begin{equation}
E^{(5)}_{A2} = \frac{8}{3\,\pi}\,\alpha^2\,(Z\,\alpha)^3\,\vec{\mu}\cdot\vec{B}\left[ \ln\frac{2\,\epsilon}{m\,(Z\,\alpha)^2}- \ln k_3-\frac{1}{3}\right].
\label{eq:e5l2}
\end{equation}
Dropping the terms containing $\ln\epsilon$ in $E^{(5)}_{A1}$ and $E^{(5)}_{A2}$, because they are already contained by definition in $E^{(5)}_{LB}$ in Eq.\,\eqref{eq:e5lb}, finally yields
\begin{equation}
\begin{split}
E^{(5)}_{LA}&=E^{(5)}_{A1}+E^{(5)}_{A2}\\
&=\frac{8}{9\,\pi}\,\alpha^2\,(Z\,\alpha)^3\,\vec{\mu}\cdot\vec{B}(2\ln k_0 - 3 \ln k_3).
\end{split}
\end{equation}
\begin{figure}
	\centering
	\includegraphics[width=1\linewidth]{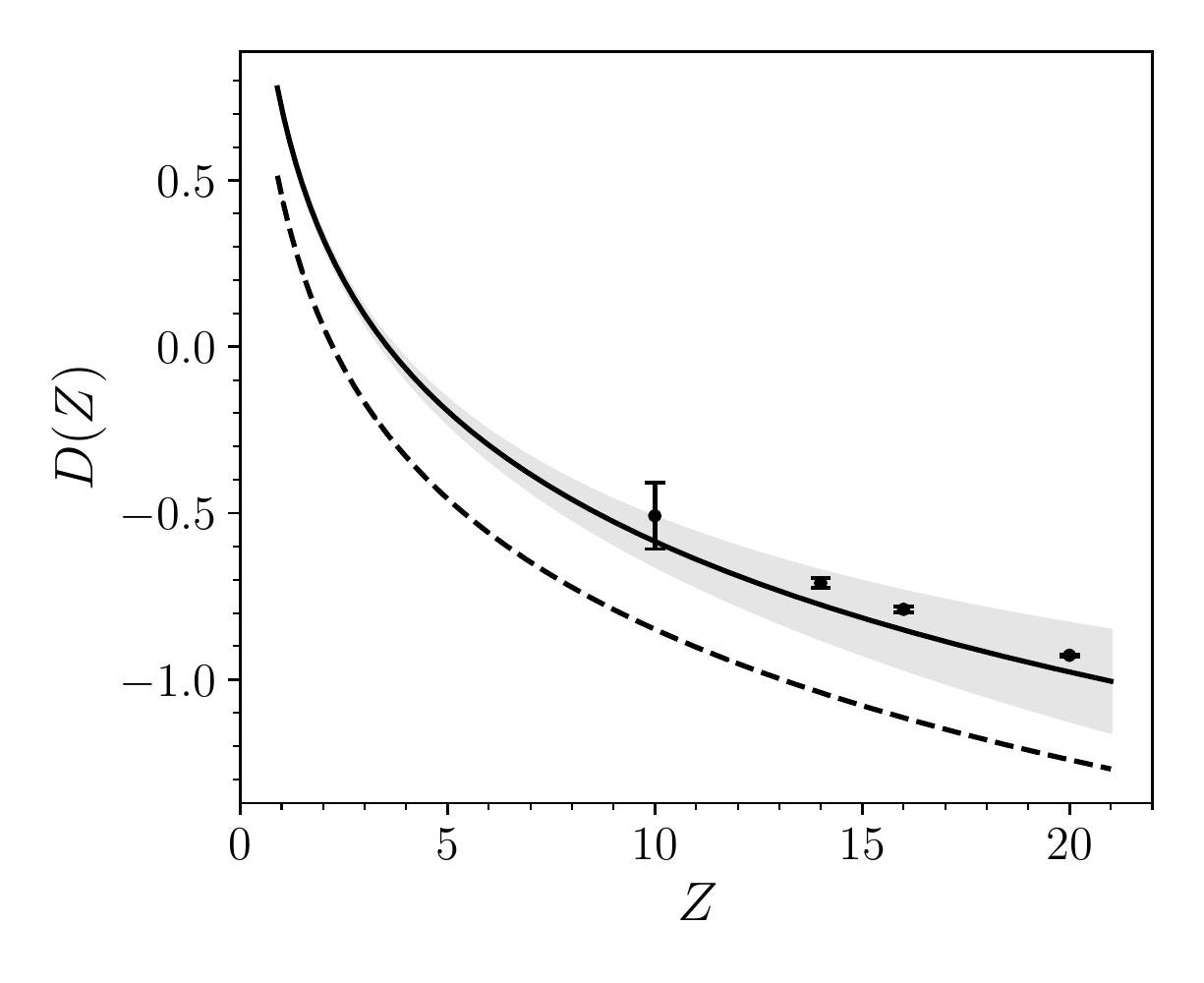}
	\caption{QED correction to the magnetic shielding expressed through the function $D(Z)$ in Eq.\,\eqref{eq:dzanal}. 
	             The solid line corresponds to the analytic result obtained in this work, and the shaded area corresponds to the estimated error of the order $Z\,\alpha$. 
	             The dots show the numerical results from Refs.\ \cite{yerokhin11,yerokhin12}, which were calculated to all orders in $Z\,\alpha$. For comparison, the previous analytic result from Refs.\ \cite{yerokhin11,yerokhin12} is shown as a dashed line.}
	\label{fig:dzhydrogen}
\end{figure}
The complete QED correction is now the sum $E^{(5)}= E^{(5)}_{LA}+E^{(5)}_{LB}+E^{(5)}_{H}$, which, expressed in terms of the shielding, gives
\begin{align}
\sigma^{(5)}={}&\frac{8}{9\,\pi}\,\alpha^2\,(Z\,\alpha)^3\,\biggl[\ln\left[(Z\,\alpha)^{-2}\right] +2\ln k_0-3\ln k_3\nonumber\\
{}&- \frac{221}{64}+\frac{3}{5}\biggr].
\label{eq:e5}
\end{align}
It is worth comparing the above result with the numerical values from Refs.\ \cite{yerokhin11,yerokhin12}, 
which were calculated to all orders in $Z\,\alpha$ but exhibited large numerical cancellations for $Z<10$. Following the procedure in Refs.\ \cite{yerokhin11,yerokhin12}, 
we define the function $D(Z)$ through
\begin{equation}
\sigma^{(5)} = \alpha^2\,(Z\,\alpha)^3\,D(Z).
\end{equation}
Using the numerical values for $\ln k_0$ and $\ln k_3 $ \cite{pachucki05},
\begin{align}
\ln k_0 &=\num{2.984128556},\\
\ln k_3 &=\num{3.272806545},
\end{align}
and omitting the vacuum polarization term of $3/5$, $D(Z)$ reads
\begin{equation}
D(Z)= \frac{8}{9\,\pi}\,\big(\ln \left[(Z\,\alpha)^{-2}\right]-\num{7.303288}\big) +O(Z\,\alpha).
\label{eq:dzanal}
\end{equation}
The term $O(Z\,\alpha)$ indicates that we estimate the error of our result to be of the order $Z\,\alpha$. 
Numerical values for $D(Z)$ calculated to all orders in $Z\,\alpha$ in Refs.\ \cite{yerokhin11,yerokhin12} are shown as dots in Fig. \ref{fig:dzhydrogen} 
together with the present analytic result of Eq.\,\eqref{eq:dzanal} (solid line). The analytic curve for $D(Z)$ from Refs.\ \cite{yerokhin11,yerokhin12} 
is shown as a dashed line for comparison. We conclude that the present analytic result is in good agreement with the numerical calculations.
The small differences are explained by the fact that our approach is valid only up to  $O(Z\,\alpha)$. 

\subsection{Recoil correction}

Contributions to the magnetic shielding due to the finite nuclear mass have been derived in Ref. \cite{pachucki08}. 
The nonrelativistic recoil correction is known to all orders in $m/m_N$, and the lowest-order terms relevant for this work are given by
\begin{align}
\sigma^{(2,1)} =&\
\frac{Z\,\alpha^2}{3}\,\biggl( \frac{1-g_N}{g_N}  - 1\biggr)\frac{m}{m_N},\\
\sigma^{(2,2)} =&\ \frac{Z\,\alpha^2}{3}\left(1+Z\,\frac{2+3\,g_N}{2\,g_N}-2\,\frac{1-g_N}{g_N}\right)\frac{m^2}{m_N^2},
\end{align}
where the nuclear $g$-factor is defined as
\begin{equation}
g_N=\frac{m_N}{Z\,m_p}\,\frac{\mu}{\mu_N}\,\frac{1}{I}.
\label{eq:nuc_gfactor}
\end{equation}
In Eq.\,\eqref{eq:nuc_gfactor}, $m_p$ is the proton mass, $\mu_N$ is the nuclear magneton, and $\mu$ and $I$ are the magnetic moment 
and the spin of the considered nucleus, respectively. We note that the nuclear $g$-factor used here is defined analogously to the electronic $g$-factor 
and is thus, in general, different from the standard definition, except for the proton. Consequently, the interaction of the nucleus with a magnetic field 
is given by $-e_N\,g_N/(2\,m_N)\,\vec{I}\cdot\vec{B}$, where $e_N$ is the charge of the nucleus.

\subsection{Total result}
The total magnetic shielding for hydrogen-like ions including contributions up to order $\alpha^5$ is (Eq. (25) of Ref. \cite{wehrli:2021}),
\begin{align}
\sigma ={}& \frac{1}{3}\,\alpha\,(Z\,\alpha)+\frac{97}{108}\,\alpha\,(Z\,\alpha)^3\nonumber\\
{}&+ \frac{8}{9\,\pi}\,\alpha^2\,(Z\,\alpha)^3\,\biggl[\ln\left[(Z\,\alpha)^{-2}\right] +2\ln k_0-3\ln k_3\nonumber\\
{}&- \frac{221}{64}+\frac{3}{5}\biggr] + \frac{Z\,\alpha^2}{3}\,\biggl( \frac{1-g_N}{g_N}  - 1\biggr)\frac{m}{m_N}\nonumber\\
&{}+\frac{Z\,\alpha^2}{3}\left(1+Z\,\frac{2+3\,g_N}{2\,g_N}-2\,\frac{1-g_N}{g_N}\right)\frac{m^2}{m_N^2}.
\end{align}
Numerical values for the cases of $^1$H and $^3$He$^+$, following Ref. \cite{wehrli:2021}, are given in Table \ref{tab:shieldingstot}.

\begin{widetext}
\section{Magnetic shielding in helium-like ions with small nuclear charge}\label{sec:heshielding}
We now go one step further and study the nuclear magnetic shielding in helium-like systems. 
The generalization of the  Breit-Pauli Hamiltonian to the two-electron system is  (see for example Ref. \cite{pachucki04}),
\begin{align}
H_\mathrm{BP}={}&\sum_{a=1}^2 H_a + H_{12},\label{breit_gen}\\
H_a={}&\frac{\vec{\pi}_a^2}{2\,m} -\frac{\vec{\pi}_a^4}{8\,m^3} -\frac{Z\,\alpha}{r_a} -\frac{e}{2\,m}\,(1+\kappa)\,\vec{\sigma}_a\cdot\vec{B}_a +\frac{e}{8\,m^3}\,\lbrace\vec{\pi}_a^2,\vec{\sigma}_a\cdot\vec{B}_a\rbrace -\frac{e^2}{2}\left(\frac{1}{4\,m^3}+\alpha_M\right)\vec{B}_a^2\nonumber\\
{}& +\frac{2\,\pi\,Z\,\alpha}{3}\left(\frac{3}{4\,m^2}+r^2_E+r^2_{\rm vp}\right)\,\delta(\vec{r_a})
+\frac{Z\,\alpha}{4\,m^2}\,(1+2\,\kappa)\,\vec{\sigma}_a\cdot\frac{\vec{r}_a\times\vec{\pi}_a}{r_a^3}\nonumber\\
{}& +\frac{e\,\kappa}{8\,m^3}\,\lbrace\vec{\pi}_a\cdot\vec{B}_a,\vec{\sigma}_a\cdot\vec{\pi}_a\rbrace +\frac{e}{6\,m}
\left(r^2_E+r^2_{\rm vp}-\frac{3\,\kappa}{4\,m^2}\right)(\nabla_a\times \vec{B}_a)\cdot\vec{\pi}_a,
\\
H_{12}={}&\frac{\alpha}{r}-\frac{4\,\pi\,\alpha}{3}\left(\frac{3}{4\,m^2}+r_E^2 + \frac{1}{2}\,r^2_{\rm vp}\right)\delta(\vec{r}) - \frac{\alpha}{2\,m^2}\,\pi_1^i\left(\frac{\delta^{ij}}{r}+\frac{r^i\,r^j}{r^3}\right)\pi_2^j  -\frac{2\,\pi\,\alpha}{3\,m^2}\,(1+\kappa)^2\,\vec{\sigma}_1\cdot\vec{\sigma}_2\,\delta(\vec{r})\nonumber\\
{}&+\frac{\alpha}{4\,m^2}\,(1+\kappa)^2\,\frac{\sigma_1^i\,\sigma_2^j}{r^3} \left(\delta^{ij}-3\,\frac{r_1^i\,r_2^j}{r^2}\right) + \frac{\alpha}{4\,m^2\,r^3}\Big[2\,(1+\kappa)\Big(\vec{\sigma}_1\cdot(\vec{r}\times\vec{\pi}_2)-\vec{\sigma}_2\cdot(\vec{r}\times\vec{\pi}_1)\Big)\nonumber\\
{}&+(1+2\,\kappa)\Big(\vec{\sigma}_2\cdot(\vec{r}\times\vec{\pi}_2)-\vec{\sigma}_1\cdot(\vec{r}\times\vec{\pi}_1)\Big)\Big],\label{eq:breithamiltonian}
\end{align}
where $\vec{r}=\vec{r}_1-\vec{r}_2$. $H_a$ corresponds to the one-electron NRQED Hamiltonian given in Eq.\,\eqref{eq:h_nrqed}, 
and the QED parameters $\kappa$, $r_E^2$, $r_\mathrm{vp}^2$, and $\alpha_M$ are given in Eqs.\,\eqref{eq:kappa}, \eqref{eq:re2}, \eqref{eq:rvp}, and \eqref{eq:alpham}, respectively. 
Note that $H_\mathrm{BP}$ can be naturally generalized to any number of electrons, and thus the generalization of QED theory of shielding
to arbitrary light atoms is more or less straightforward.
The magnetic shielding in helium has already been studied in Ref. \cite{rudzinski09}, in which
the complete result for the relativistic correction was derived, together with the  leading  logarithmic QED contribution. 
In the following sections we present the derivation and numerical calculation of the complete QED correction.

\subsection{Leading-order $\alpha^2$ and relativistic contribution $\alpha^4$}
The leading-order nuclear magnetic shielding can be directly deduced from the hydrogen-like case because it is simply the sum of the leading contribution from the two electrons,
\begin{equation}
\sigma^{(2)}=\frac{\alpha}{3\,m}\left\langle\frac{1}{r_1}+\frac{1}{r_2}\right\rangle.
\end{equation}
The expectation value is taken with respect to the ground state $|\phi_0\rangle$ of the nonrelativistic Hamiltonian $H_0$,
\begin{align}
H_0&=\frac{\vec{p}_1^{\,2}}{2\,m}+\frac{\vec{p}_2^{\,2}}{2\,m}-\frac{Z\,\alpha}{r_1}-\frac{Z\,\alpha}{r_2}+\frac{\alpha}{r}.
\end{align}
For the derivation of the relativistic correction, we start from Eq.\,\eqref{eq:ecorrectiongeneral} and neglect all the QED corrections,
\begin{align}
E^{(4)}={}&\Bigg\langle\sum_a\bigg[-\frac{e^2}{4\,m^3}\,\lbrace \vec{p}_a^{\,2},\vec{A}_{E,a}\cdot\vec{A}_{I,a}\rbrace -\frac{e^2}{4\,m^3}\,\vec{B}_{I,a} \cdot \vec{B}-\frac{e^2}{2\,m^3}\,\lbrace\vec{p}_a\cdot \vec{A}_{E,a},\vec{p}_a\cdot\vec{A}_{I,a}\rbrace\bigg]\nonumber\\
{}&-\frac{e^2\,\alpha}{2\,m^2}\left(\frac{\delta^{ij}}{r} + \frac{r^i\,r^j}{r^3}\right)(A_{I,1}^iA_{E,2}^j+A_{E,1}^iA_{I,2}^j)\Bigg\rangle\nonumber\\
{}&+ 2\,\Bigg\langle\sum_a\left[\frac{e^2}{m}\,\vec{A}_{E,a}\cdot\vec{A}_{I,a}\right] \frac{1}{(E_0-H_0)'} \bigg \lbrace\sum_a\bigg[-\frac{\vec{p}_a^{\,4}}{8\,m^3}+\frac{\pi\, Z\,\alpha}{2\,m^2}\,\delta(\vec{r}_a)\bigg]-\frac{\pi\,\alpha}{m^2}\,\delta(\vec{r})\nonumber\\
{}&  - \frac{\alpha}{2\,m^2}\,p_1^i\left(\frac{\delta^{ij}}{r}+\frac{r^i\,r^j}{r^3}\right)p_2^j-\frac{2\,\pi\,\alpha}{3\,m^2}\,\vec{\sigma}_1\cdot\vec{\sigma}_2\,\delta(\vec{r}) \bigg\rbrace\Bigg\rangle\nonumber\\
{}& + 2\,\Bigg\langle\sum_a\left[-\frac{e}{m}\,\vec{p}_a\cdot\vec{A}_{I,a}-\frac{e}{2\,m}\,\vec{\sigma}_a\cdot\vec{B}_{I,a} \right] \frac{1}{(E_0-H_0)'}\bigg\lbrace\sum_a\bigg[\frac{e}{4\,m^3}\,\lbrace \vec{p}_a^{\,2},\vec{p}_a\cdot\vec{A}_{E,a}\rbrace\nonumber\\
{}& + \frac{e}{8\,m^3}\,\lbrace \vec{p}_a^{\,2},\vec{\sigma}_a\cdot\vec{B}\rbrace  - \frac{e\,Z\,\alpha}{4\, m^2}\,\vec{\sigma}_a\cdot\left(\frac{\vec{r}_a}{r_a^3}\times\vec{A}_{E,a}\right) \bigg] +\frac{e\,\alpha}{2\,m^2}\left(\frac{\delta^{ij}}{r}+\frac{r^i\,r^j}{r^3}\right)(A_{E,1}^ip_2^j + p_1^iA_{E,2}^j)\nonumber\\
{}&-\frac{e\,\alpha}{4\,m^2\,r^3}\,\Big[(2\,\vec{\sigma}_1+\vec{\sigma}_2)\cdot(\vec{r}\times\vec{A}_{E,2})-(\vec{\sigma}_1 + 2\,\vec{\sigma}_2)\cdot(\vec{r}\times\vec{A}_{E,1})\Big] \bigg\rbrace\Bigg\rangle,
\label{eq:e4generalhelium}
\end{align}
where $\sum_a$ denotes a sum over $a=1,2$.
Because the helium ground state is a singlet state, we have for the expectation value of spin-spin terms $\langle\sigma_1^i\,\sigma_2^j\rangle = -\delta^{ij}$;
thus, $\sigma^{(4)}$ becomes
\begin{align}
\sigma^{(4)}={}&\frac{\alpha^2}{12\,m^2}\left\langle\left[\frac{1}{r_1^3}+\frac{1}{r_2^3}\right]\left(\frac{(\vec{r}\cdot\vec{r}_1)(\vec{r}\cdot\vec{r}_2)}{r^3}-3\,\frac{\vec{r}_1\cdot\vec{r}_2}{r}\right)\right\rangle-\frac{\alpha}{6\,m^3}\sum_a\left\langle\frac{1}{r_a}\,\vec{p}_a^{\,2}+\frac{\vec{L}_a^2}{r_a^3}+4\,\pi\,\delta(\vec{r}_a)\right\rangle\nonumber\\
{}&+\frac{2\,\alpha}{3\,m^3}\left\langle\left[\frac{1}{r_1}+\frac{1}{r_2}\right]\frac{1}{(E_0-H_0)'}\left[\sum_a\left(\frac{\pi\, Z\,\alpha}{2}\,\delta(\vec{r}_a)-\frac{\vec{p}^{\,4}_a}{8\,m}\right)+\pi\,\alpha\,\delta(\vec{r}) -\frac{\alpha}{2}\,p_1^i\left(\frac{\delta^{ij}}{r}+\frac{r^i\,r^j}{r^3}\right)p_2^j\right]\right\rangle\nonumber\\
{}&-\frac{\alpha}{6\,m^3}\bigg\langle\left(\frac{\vec{L}_1}{r_1^3}+\frac{\vec{L}_2}{r_2^3}\right)\frac{1}{E_0-H_0}\,\bigg[\frac{1}{m}\left(\vec{p}_1^{\,2}\,\vec{L}_1+\vec{p}_2^{\,2}\,\vec{L}_2\right)+\alpha\,\bigg(\frac{1}{r}\,\vec{r}_1\times\vec{p}_2+\frac{1}{r}\,\vec{r}_2\times\vec{p}_1-\vec{r}_1\times\vec{r}_2\,\biggl[\frac{\vec{r}}{r^3}\cdot(\vec{p}_1+\vec{p}_2)\biggr]\bigg)\bigg]\bigg\rangle\nonumber\\
{}&-\frac{2\,\alpha}{3\,m^3}\left\langle\pi\left(\delta(\vec{r}_1)-\delta(\vec{r}_2)\right)\frac{1}{E_0-H_0} \left[\frac{\vec{p}_1^{\,2}}{m}-\frac{\vec{p}_2^{\,2}}{m} -\frac{Z\,\alpha}{3\,r_1}+\frac{Z\,\alpha}{3\,r_2}-\frac{\alpha}{3\,r^3}\,\vec{r}\cdot(\vec{r}_1+\vec{r}_2)\right]\right\rangle\nonumber\\
{}&-\frac{\alpha^2}{8\,m^3}\left\langle\left(\frac{r_1^i\,r_1^j}{r_1^5}-\frac{r_2^i\,r_2^j}{r_2^5}\right)^{(2)}\frac{1}{E_0-H_0}\left(Z\,\frac{r_1^i\,r_1^j}{r_1^3}-Z\,\frac{r_2^i\,r_2^j}{r_2^3}+\frac{r^i\,(r_1^j+r_2^j)}{r^3}\right)^{(2)}\right\rangle.
\end{align}
This result reduces to the hydrogenic case, Eq.\,\eqref{eq:sigma4hydrogen}, after removing the terms that involve the second electron.

\subsection{$\alpha^5$ QED correction in He without magnetic fields}\label{sec:helambshift}
We rederive here the helium Lamb shift following closely the former work \cite{pachucki98}, and for this we use the generalized Breit-Pauli Hamiltonian in Eq.\,\eqref{breit_gen},
which accounts for most of the QED effects. The helium Lamb shift is  split into the radiative correction, 
in which a photon is emitted and absorbed by the same electron, and the exchange one, where a photon is exchanged between the two electrons.

The radiative correction is split into the low- and high-energy parts $E^{(5)}_{SL} + E^{(5)}_{SH}$.
The low-energy part in the electron self-energy $E_{SL}$ is (see also Eq.\,\eqref{eq:ellamb})
\begin{equation}
E^{(5)}_{SL} = \frac{2\,\alpha^2}{3\,\pi\,m^2}\,\int_0^{\epsilon}
dk\,k
\left\langle\vec{p}_1\,\frac{1}{E_0-k-H_0}\,\vec{p}_1\,
\right\rangle+(1\rightarrow 2),
\end{equation}
where the term $(1\rightarrow 2)$ is identical to the first one but with electronic indices 2 instead of 1. This expression is linearly divergent in $\epsilon$. 
Because after expansion in $\alpha$, one goes with $\epsilon$ to 0, the linear term can be subtracted out, but the logarithmic term stays.
After $k$--integration, $E^{(5)}_{SL}$ takes the form
\begin{align}
E^{(5)}_{SL} =&{} \frac{2\,\alpha}{3\,\pi\,m^2}\, 
\left\langle \vec{p}_1\,(H_0-E_0)\,
\ln\left(\frac{\epsilon}{H_0-E_0}\right)
\vec{p}_1\right\rangle +(1\rightarrow 2)\nonumber\\
=&{}\left[-\frac{2\,\alpha}{3\,\pi\,m^2}
\left\langle \vec{p}_1\,(H_0-E_0)\,
\ln\left(\frac{2\,(H_0-E_0)}{m\,\alpha^2}\right)
\vec{p}_1\right\rangle +(1\rightarrow 2)\right]
+\left[\frac{4\,Z\,\alpha^2}{3\,m^2}\,\left\langle
\delta(\vec{r}_1)+\delta(\vec{r}_2)\right\rangle
-\frac{8\,\alpha^2}{3\,m^2}\,\langle\delta(\vec{r})\rangle
\right]\,\ln\frac{2\,\epsilon}{m\,\alpha^2}\,.
\end{align}
When combined with the low-energy part from the photon exchange,
it will form a Bethe logarithm for the helium atom. 

The high-energy part $E^{(5)}_{SH}$ is obtained from the generalized Breit-Pauli Hamiltonian
in a similar way as for hydrogenic systems, namely
\begin{align}
E^{(5)}_{SH} =&\ 
\sum_a \frac{2\,\pi\,Z\,\alpha}{3}\,(r_E^2+r_{\rm vp}^2)\,\langle\delta(\vec{r_a})\rangle -\frac{4\,\pi\,\alpha}{3}\,\biggl(r_E^2 + \frac{1}{2}\,r_{\rm vp}^2 
+ \frac{\kappa}{m^2}\,\vec\sigma_1\cdot\vec\sigma_2\biggr)\,\langle\delta(\vec{r})\rangle
\nonumber \\ =&\
 \frac{Z\,\alpha^2}{m^2}\,\left(
\frac{10}{9}-\frac{4}{15}-\frac{4}{3}\,\ln{2\,\epsilon}\right)\,
\langle\delta(\vec{r}_1)+\delta(\vec{r}_2)\rangle
-\frac{\alpha^2}{m^2}\,\left( 2\cdot\frac{10}{9}-\frac{4}{15}-2\cdot\frac{4}{3}\,\ln{2\,\epsilon} -2 \right)\,\langle\delta(\vec{r})\rangle,
\end{align}
where $10/9$ comes from the electron self-energy and $4/15$
from the vacuum polarization, and the last term $(-2)$ is from the spin-spin interaction. 

We now move on to the exchange terms. The single transverse photon exchange diagrams lead to the correction
\begin{align}
E^{(5)}_T = &\ \frac{e^2}{m^2}\,\int\frac{d^3k}{(2\,\pi)^3\,2\,k}\,
\left(\delta^{ij}-\frac{k^i\,k^j}{k^2}\right)\,
\left\langle p_1^i\,e^{i\,\vec k\cdot\vec r_1}\,\frac{1}{E_0-k-H_0}\,p_2^j\,e^{-i\,\vec k\cdot\vec r_2}\right\rangle+\mbox{h.c.}\,,
\label{9}
\end{align}
which is split into the low- and middle-energy parts $E^{(5)}_T = E^{(5)}_L + E^{(5)}_M$. The low-energy part $E^{(5)}_L$ reads
\begin{align}
E^{(5)}_L ={}& \frac{2\,e^2}{3\,m^2}\,\int_{k<\epsilon}
\frac{ d^3k}{(2\,\pi)^3\,2\,k}\,
\left\langle\, \vec{p}_1\,\frac{1}{E_0-H_0-k}\,\vec{p}_2\,
\right\rangle+\mbox{h.c.} \nonumber \\ 
={}&
\frac{2\,\alpha}{3\,\pi\,m^2}\, 
\left\langle \vec{p}_1\,(H_0-E_0)\,
\ln\left(\frac{\epsilon}{H_0-E_0}\right)
\vec{p}_2\right\rangle+(1\leftrightarrow 2)\nonumber \\ 
={}&
\left[-\frac{2\,\alpha}{3\,\pi\,m^2}\, 
\left\langle \vec{p}_1\,(H_0-E_0)\,
\ln\left(\frac{2(H_0-E_0)}{m\,\alpha^2}\right)
\vec{p}_2\right\rangle+(1\leftrightarrow 2)\right] 
+\frac{8}{3}\,\frac{\alpha^2}{m^2}\,\ln\frac{2\,\epsilon}{m\,\alpha^2}\,
\langle\delta(\vec{r})\rangle \,.
\end{align}
The middle-energy part is obtained from Eq.\,\eqref{9} with the condition 
$k>\epsilon$, which allows us to perform the following expansion of the denominator in Eq.\,\eqref{9}
\begin{equation}
\frac{1}{E_0-H_0-k} = -\frac{1}{k}+\frac{H_0-E_0}{k^2}+\ldots
\end{equation}
The first term gives an energy correction of order $m\,\alpha^4$ and is already included in the Breit Hamiltonian.
The next term contributes at $m\,\alpha^5$. The following terms denoted by ellipses are of higher order and are neglected. Due to this expansion, the middle-energy part becomes
\begin{align}
E^{(5)}_M = \frac{e^2}{m^2}\,\int_{k>\epsilon}
\frac{d^3k}{(2\,\pi)^3\,2\,k^3}\,
\left(\delta^{ij}-\frac{k^i\,k^j}{k^2}\right)\,
\left\langle p_1^i\, e^{i\,\vec{k}\cdot\vec{r}_1}\,(H_0-E_0)\,p_2^j\, e^{-i\,\vec{k}\cdot \vec{r}_2}\right\rangle+\mbox{h.c.}
\end{align}
The matrix element can be further rewritten as
$(\vec{r}=\vec{r}_1-\vec{r}_2)$,
\begin{align}
\langle\cdots\rangle+\mbox{h.c.} ={}&
\left\langle \Bigl[ p_1^i\, e^{i\,\vec{k}\cdot\vec{r}_1}\,,\Bigl[H_0-E_0\,,\, p_2^j\, e^{-i\,\vec{k}\cdot \vec{r}_2}\Bigr]\Bigr]\right\rangle  
=
\left\langle e^{i\,\vec{k}\cdot\vec{r}}\, \left[p_1^i\,,\left[
\frac{\alpha}{r}\,,p_2^j\right]\right]\right\rangle  
\end{align}
and the middle-energy part becomes
\begin{equation}
E^{(5)}_M = \frac{e^2}{m^2}\,\int_{k>\epsilon}
\frac{ d^3k}{(2\,\pi)^3\,2\,k^3}\,
\left(\delta^{ij}-\frac{k^i\,k^j}{k^2}\right)\,
\left\langle e^{i\,\vec{k}\cdot\vec{r}} \left[p^i\,,\left[p^j\,,
\frac{\alpha}{r}\right]\right]\right\rangle\,. \label{18} 
\end{equation}
After $k$-integration we obtain
\begin{equation}
E^{(5)}_M = -\frac{2\,m\,\alpha^5}{3\,\pi}\,
\left\langle P\left(\frac{1}{(m\,\alpha\,r)^3}\right)\right\rangle
+\frac{8}{3}\,\frac{\alpha^2}{m^2}\,
\left(\frac{4}{3}-\ln\frac{\epsilon}{m\,\alpha}\right)\,\langle\delta(\vec{r})\rangle\,,
\end{equation}
where
\begin{equation}
\langle\phi|P\left(\frac{1}{r^3}\right)|\psi\rangle=\lim_{a\rightarrow0}\int d^3r\,\phi^*(\vec{r})\left[\frac{1}{r^3}\,\Theta(r-a)+4\,\pi\,\delta(\vec{r})\,(\gamma+\ln a)\right]\psi(\vec{r}).
\label{eq:pdistribution}
\end{equation}
The double transverse photon exchange, namely the double seagull $E^{(5)}_S$ and the hard two-photon exchange $E^{(5)}_H$, are considered together
because they are not affected by the presence of the magnetic field,
\begin{equation}
E^{(5)}_S+E^{(5)}_H = -\frac{m\,\alpha^5}{2\,\pi}\,
\left\langle P\left(\frac{1}{(m\,\alpha\,r)^3}\right)\right\rangle
-\frac{\alpha^2}{m^2}\,
\left(\frac{8}{3}\,\ln2-\frac{22}{3}-2\,\ln\alpha\right)
\,\langle\delta(\vec{r})\rangle\,.
\end{equation}

The complete helium Lamb shift $E^{(5)}$ is the sum of the terms $E^{(5)}_{SL}+E^{(5)}_{SH}+E^{(5)}_{L}+E^{(5)}_{M}+E^{(5)}_{S}+E^{(5)}_{H}$,
\begin{align}
E^{(5)} ={}& \left[\frac{164}{15}+\frac{14}{3}\,\ln\alpha\right]\,\frac{\alpha^2}{m^2}\,
\langle\delta(\vec{r})\rangle +\left[\frac{19}{30}+\ln(\alpha^{-2})\right]\,
\frac{4\,\alpha^2\,Z}{3\,m^2}\,\langle\delta(\vec{r}_1)+\delta(\vec{r}_2)\rangle
\nonumber \\{}&
-\frac{14}{3}\,m\,\alpha^5\,\left\langle
\frac{1}{4\,\pi}\,P\left(\frac{1}{(m\,\alpha\,r)^3}\right)\right\rangle-
\frac{2\,\alpha}{3\,\pi\,m^2}\,{\cal M}\,, \label{eq:helambshift}
\end{align}
where
\begin{align}
{\cal M} 
={}&\left\langle(\vec{p}_1+\vec{p}_2)\,(H_0-E_0)\,
\ln\frac{2\,(H_0-E_0)}{m\,\alpha^2}\,
(\vec{p}_1+\vec{p}_2)\right\rangle 
= 2\,\pi\,\alpha\,Z\,\langle\delta(\vec{r}_1)+\delta(\vec{r}_2)\rangle\,\ln k_0\,,
\end{align}
and where $\ln k_0$ is the Bethe logarithm for the helium atom.

\subsection{$\alpha^5$ QED correction to the magnetic shielding in He}
We derive here the leading QED correction to the shielding, bearing in mind the derivation of the QED correction to energy from the previous section. 
We therefore split the QED correction $\sigma^{(5)}$ as
\begin{equation}
\sigma^{(5)}=\sigma_{B0} + \sigma_{H}  + \sigma_{L},
\label{eq:hee5sum}
\end{equation}
where $\sigma_{B0}$ is the Lamb shift $E^{(5)}$ with the wave function corrected by the leading shielding,
$\sigma_H$ is the high-energy part beyond $\sigma_{B0}$, and $\sigma_L$ is the low-energy part.
The correction to the wave function is (see Eqs. \eqref{eq:phiapprox} and \eqref{eq:helambshift})
\begin{align}
\sigma_{B0} ={}& \frac{2\,\alpha^3}{3\,m^3}\,\biggl\langle\biggl(\frac{1}{r_1}+\frac{1}{r_2}\biggr)\frac{1}{(E_0-H_0)'}\biggl\{
\left(\frac{164}{15}+\frac{14}{3}\,\ln\alpha\right)\delta(\vec{r})\nonumber\\ 
{}&+\left(\frac{5}{6}-\frac{1}{5}+\ln(\alpha^{-2})\right)\frac{4\,Z}{3}\,[\delta(\vec{r_1})+\delta(\vec{r_2})]
-\frac{7\,\alpha^3\,m^3}{6\,\pi}\,P\left(\frac{1}{(m\,\alpha\, r)^3}\right)\biggr\}\biggr\rangle.
\end{align}
The high-energy part $E_H$ is directly obtained from the generalized Breit-Pauli Hamiltonian
\begin{align}
E_{H} =&\ \sum_a\biggl\langle -e^2\,\alpha_M\,\vec{B}_a\cdot\vec{B}_{I,a}
  -\frac{e^2}{6m} \left(r^2_E+r^2_{\rm vp}-\frac{3\,\kappa}{4\,m^2}\right)\, \vec{A}_{E,a}\cdot(\vec\nabla_a\times \vec{B}_{I,a})\bigg\rangle \nonumber\\
  &\ +2\,\bigg\langle -\frac{e\,\kappa}{2\,m}\,\sum_b\vec\sigma_b\cdot\vec B_{I,b}\,\frac{1}{(E_0-H_0)'}\,\biggl\{\sum_{a}\biggl[
    \frac{e}{8\,m^3}\,\{p_a^2\,,\,\vec\sigma_a\cdot\vec B\}
    +\frac{e}{8\,m^3}\,\{\vec p_a\cdot\vec B\,,\,\vec\sigma_a\cdot\vec p_a\}
    \nonumber \\ &\
  -\frac{3\,e\,Z\,\alpha}{4\,m^2}\,\vec\sigma_a\cdot\left(\frac{\vec r_a}{r_a^3}\times\vec A_{E,a}\right)\biggr]
  -\frac{e\,\alpha}{4\,m^2\,r^3}\Big[(2\,\vec{\sigma}_1+\vec{\sigma}_2)\cdot(\vec{r}\times\vec{A}_{E,2})
  -(2\,\vec{\sigma}_2+\vec{\sigma}_1)\cdot(\vec{r}\times\vec{A}_{E,1})\Big]\bigg\} \bigg\rangle.
 \end{align}
This can be expressed as the following correction to the shielding constant                             
    \begin{align}
   \sigma_{H}   =&\ \sigma'_{B1} +\sigma_{B2} + \sigma_{B3},\nonumber\\ 
  \sigma'_{B1} =&\    \frac{\alpha^2}{m^3}\biggl(\frac{20}{9}\,\ln\frac{m}{2\,\epsilon}-\frac{301}{108}+\frac{4}{15}\biggr)\left\langle\delta(\vec{r}_1) + \delta(\vec{r}_2)\right\rangle,\nonumber  \\
  \sigma_{B2} =&\ 
  -\frac{\alpha^2}{3\,m^3}\left\langle\left(\delta(\vec{r}_1)-\delta(\vec{r}_2)\right)\frac{1}{E_0-H_0}
  \left[\frac{4\,\vec{p}_1^2}{3\,m}-\frac{4\,\vec{p}_2^2}{3\,m}-\frac{Z\,\alpha}{r_1}+\frac{Z\,\alpha}{r_2}-\frac{\alpha}{3\,r^3}\,\vec{r}\cdot(\vec{r}_1+\vec{r}_2)\right]\right\rangle, \nonumber \\ 
  \sigma_{B3} =&\
  -\frac{3\,\alpha^2}{16\,\pi\, m^3}\left\langle\left(\frac{r_1^i\,r_1^j}{r_1^5}-\frac{r_2^i\,r_2^j}{r_2^5}\right)^{(2)}\!\!\!\frac{1}{E_0-H_0}
  \left(Z\,\alpha\,\frac{r_1^i\,r_1^j}{r_1^3}-Z\,\alpha\,\frac{r_2^i\,r_2^j}{r_2^3}+\alpha\,\frac{r^i\,(r_1^j+r_2^j)}{3\,r^3}+\frac{2}{3\,m}\,(p_1^i\,p_1^j-p_2^i\,p_2^j)\right)^{(2)}\right\rangle.
\end{align}
We now turn to the low-energy part. Analogously to Eqs.\,\eqref{eq:e5lstart}\,--\,\eqref{eq:e5lb}, we have
\begin{equation}
  E_L =\frac{2\,\alpha}{3\,\pi\, m^2}\int_{0}^{\epsilon}dk\,k\left\langle(\vec{\pi}_1+\vec{\pi}_2)\,\frac{1}{E-H-k}\,(\vec{\pi}_1+\vec{\pi}_2)\right\rangle_{\!B} = E_{LA} + E_{LB},
\end{equation}
where $\langle\ldots\rangle_{B}$ denotes the expectation value with respect to the ground state with energy $E$ of the Hamiltonian
\begin{equation}
H=\frac{\vec{\pi}^2_1}{2\,m}+\frac{\vec{\pi}^2_2}{2\,m}-\frac{Z\,\alpha}{r_1}-\frac{Z\,\alpha}{r_2}+\frac{\alpha}{r},
\end{equation}
and where
\begin{align}
E_{LA} &= -\frac{2\,\alpha}{3\,\pi\,m^2}\left\langle(\vec \pi_1+\vec\pi_2)\,(H-E)\ln\frac{2\,(H-E)}{m\,\alpha^2}\,(\vec\pi_1+\vec\pi_2)\right\rangle_{\!B},\label{eq:hee5la}\\
E_{LB}&=\frac{\alpha}{3\,\pi\, m^2}\ln\frac{2\,\epsilon}{m\,\alpha^2}\left\langle[\vec{\pi}_1+\vec{\pi}_2,[H-E,\vec{\pi}_1+\vec{\pi}_2]]\right\rangle_{B} \nonumber\\
&=\vec{\mu}\cdot\vec{B}\,\frac{\alpha^2}{m^3}\ln\frac{2\,\epsilon}{m\,\alpha^2}\left[\frac{20}{9}\left\langle\delta(\vec{r}_1)+\delta(\vec{r}_2)\right\rangle+\frac{8\,Z\,\alpha}{9}\left\langle\left(\frac{1}{r_1}+\frac{1}{r_2}\right)\frac{1}{(E_0-H_0)'}\left(\delta(\vec{r}_1)+\delta(\vec{r}_2)\right)\right\rangle\right].
\end{align}
We thus obtain
\begin{align}
\sigma_{LB} =&\  \ln\frac{2\,\epsilon}{m\,\alpha^2}\frac{20}{9}\left\langle\delta(\vec{r}_1)+\delta(\vec{r}_2)\right\rangle, \label{eq:heelb}
\end{align}
in which we have dropped the terms already included in $\sigma_{B0}$.
We observe that the divergent terms in Eq.\,\eqref{eq:heelb} cancel with those in $\sigma_{H}$,
\begin{align}
\sigma_{B1} =&\ \sigma'_{B1}  + \sigma_{LB},
\end{align}
as expected.

Equation\,\eqref{eq:hee5la} is only a formal expression for $E_{LA}$, and it needs to be expanded in the magnetic field. 
For this we rewrite $E_{LA}$ in the form
\begin{align}
  E_{LA} =&\ -\frac{2\,\alpha}{3\,\pi}\left\langle(\vec r_1+\vec r_2)\,(H-E)^3\ln\frac{2\,(H-E)}{m\,\alpha^2}\,(\vec r_1+\vec r_2)\right\rangle_{\!B},
\end{align}
and the Hamiltonian $H$ as
\begin{align}
  H&=H_0+\frac{1}{3\,m}\,\vec{\mu}\cdot\vec{B}\,U  -\frac{e}{2\,m}\,\vec{L}\cdot\vec{B} -\frac{e}{4\,\pi\, m}\vec\mu\cdot \vec U, \label{98}
\end{align}
where $L^i =L_1^i+L_2^i$, $U = \alpha/r_1 + \alpha/r_2$, and $U^i =L_1^i/r_1^3+L_2^i/r_2^3$.
We note that $L^i|\phi_0\rangle=0$ and $\langle U^i\rangle=0$.
The fact that $U^i|\phi_0\rangle\neq0$, in contrast to the hydrogenic case, makes the evaluation of $E_{LA}$ more complicated for the helium atom.
Following the hydrogenic case, $E_{LA}$ is  split again into two parts
\begin{align}
  E_{LA} =&  E_{A1} +  E_{A2},
\end{align}
where each part comes from different perturbations.
The first part due to $1/(3\,m)\,\vec{\mu}\cdot\vec{B}\, U$ is given by
\begin{align}
  \sigma_{A1} =&\ -\frac{2\,\alpha}{9\,\pi}\,\delta_{U}\left\langle(\vec r_1+\vec r_2)\,(H_0-E_0)^3\ln\frac{2\,(H_0-E_0)}{m\,\alpha^2}\,(\vec r_1+\vec r_2)\right\rangle
  \nonumber  \\ =&\
  \frac{2\,\alpha}{9\,\pi\, m^2}\,{\cal D}_{A1}\,\ln k_{A1},
\end{align}
where 
\begin{align}
  {\cal D}_{A1} \equiv&\ -4\, \pi\,Z\left\langle U\, \frac{1}{(E_0-H_0)'}\, (\delta(\vec{r_1})+\delta(\vec{r_2})) \right\rangle  
  + 2\, \pi\,\alpha\left\langle \delta(\vec{r_1})+\delta(\vec{r_2}) \right\rangle. \label{101}
\end{align}
A detailed derivation of $\ln k_{A1}$ is presented in Appendix A.
The second part is due to perturbation from the two other terms in Eq.\,\eqref{98},
\begin{align}
  \sigma_{A2} =&\ -\frac{\alpha^2}{9\,\pi\,m^2}\,\delta_{L^i}\delta_{U^i}
  \left\langle(\vec r_1+\vec r_2)\,(H_0-E_0)^3\ln\frac{2\,(H_0-E_0)}{m\,\alpha^2}\,(\vec r_1+\vec r_2)\right\rangle\\
  =&\ 
   -\frac{\alpha^2}{9\,\pi\,m^2}\,(1+3\,\ln k_{A2})\, {\cal D}_{A2},
\end{align}
where
\begin{align}  
 {\cal D}_{A2} =&\   8\,\pi\left\langle \delta(\vec{r_1})+\delta(\vec{r_2}) \right\rangle. \label{104}
 \end{align}
A detailed derivation of $\ln k_{A2}$ is also presented in Appendix A.

The final result in atomic units and using the notation $\sigma^{(n)}=\alpha^n\tilde{\sigma}^{(n)}$ is
\begin{align}
  \tilde\sigma^{(5)}={}&  \tilde\sigma_{B0} + \  \tilde\sigma_{B1} +  \tilde\sigma_{B2} +  \tilde\sigma_{B3} +  \tilde\sigma_{A1} +  \tilde\sigma_{A2},\\
 \tilde\sigma_{B0}={}&\frac{2}{3}\,\biggl\langle\biggl(\frac{1}{r_1}+\frac{1}{r_2}\biggr)\frac{1}{(E_0-H_0)'}\biggl[\frac{4\,Z}{3}\left(\frac{19}{30}+\ln(\alpha^{-2})\right)[\delta(\vec{r_1})+\delta(\vec{r_2})] + 
\left(\frac{164}{15}+\frac{14}{3}\,\ln\alpha\right)\delta(\vec{r}) -\frac{7}{6\,\pi}\, P\left(\frac{1}{r^3}\right)\biggr]\biggr\rangle, \\   
\tilde\sigma_{B1}={}&\left(\frac{20}{9}\,\ln(\alpha^{-2})-\frac{1361}{540}\right)\left\langle\delta(\vec{r}_1)+\delta(\vec{r}_2)\right\rangle,  \\
\tilde\sigma_{B2}={}& -\frac{1}{3}\left\langle\left(\delta(\vec{r}_1)-\delta(\vec{r}_2)\right)\frac{1}{E_0-H_0}\left[\frac{4}{3}\,\vec{p}_1^{\,2}-\frac{4}{3}\,\vec{p}_2^{\,2}                                   -\frac{Z}{r_1}+\frac{Z}{r_2}-\frac{1}{3\,r^3}\,\vec{r}\cdot(\vec{r}_1+\vec{r}_2)\right]\right\rangle \\
 \tilde\sigma_{B3}={}& -\frac{3}{16\,\pi}\left\langle\left(\frac{r_1^i\,r_1^j}{r_1^5}-\frac{r_2^i\,r_2^j}{r_2^5}\right)^{(2)}\frac{1}{E_0-H_0}                              \left[Z\,\frac{r_1^i\,r_1^j}{r_1^3}-Z\,\frac{r_2^i\,r_2^j}{r_2^3}+\frac{r^i\,(r_1^j+r_2^j)}{3\,r^3}+\frac{2}{3}\,(p_1^i\,p_1^j-p_2^i\,p_2^j)\right]^{(2)}\right\rangle, \\
 \tilde\sigma_{A1}={}& \frac{2}{9\,\pi}\,{\cal D}_{A1}\,\ln k_{A1},\\
 \tilde\sigma_{A2}={}& -\frac{1}{9\,\pi}\,{\cal D}_{A2}\,(1+3\,\ln k_{A2}). 
\end{align}
This result for $\sigma^{(5)}$ forms the complete expression for the QED corrections of order $\alpha^5$ in the infinite nuclear mass limit for helium-like ions.

\subsection{Recoil correction}\label{sec:recoil_he}
Contributions to the magnetic shielding in helium due to the finite nuclear mass are given by \cite{pachucki08,rudzinski09}
\begin{align}
\sigma^{(2,1)} =&\ \frac{\alpha}{3\,m}\,\biggl\langle\left(\frac{1}{r_{1}}+\frac{1}{r_{2}}\right)
\frac{1}{(E_0-H_0)'}\,\frac{\vec{p}_N^{\,2}}{m_N}\biggr\rangle
+\frac{1}{3\,m}\,\frac{(1-g_N)}{Z\,g_N}\left\langle \frac{\vec{p}_N^{\,2}}{m_N}\right\rangle\nonumber\\ 
&\ +\frac{\alpha}{3\,m}\,\biggl\langle
\bigl(\vec r_1\times\vec p_2 +\vec r_2\times\vec p_1\bigr)\,
\frac{1}{E_0-H_0}\,
\sum_{a}\,\frac{\vec r_{a}}{r_a^3}\times \frac{\vec {p}_a}{m_N}\biggr\rangle,
\label{21}
\end{align}
where $\vec{p}_N=-\vec{p}_1-\vec{p}_2$ is the momentum of the nucleus and $g_N$ is the nuclear $g$-factor defined in Eq.\,\eqref{eq:nuc_gfactor}.
\end{widetext}

\section{Summary}
We have presented the derivation of the leading quantum electrodynamics corrections to the magnetic shielding for hydrogen- and helium-like atomic systems.
Our results for hydrogen-like ions are in a good agreement with the direct numerical calculations of Refs. \cite{yerokhin11,yerokhin12},
and we note significant cancellations for the low-$Z$ systems like $^1$H and $^3$He$^+$.
Similar numerical cancellation of the QED correction is present also for $^3$He, as shown in Table \ref{tab:numvalueshe}. One may therefore conclude
that QED effects to the nuclear magnetic shielding are not significant for low-$Z$ systems. 
\begin{table}[ht]
\caption{Numerical values for QED contributions to the nuclear magnetic shielding in $^3$He (from Ref. \cite{wehrli:2021}), which illustrate significant cancellations. All values are in atomic units with $\sigma^{(n)}=\alpha^n\tilde{\sigma}^{(n)}$.}\label{tab:numvalueshe}
\begin{center}
\begin{tabular}{ld{3.22}}
    \hline \hline \\
contribution&  \centt{value} \\[1ex]
\hline
$\tilde\sigma_{A1} $ &  33.750\,67(2) \\
$\tilde\sigma_{A2} $ & -48.007\,69(14) \\
$\tilde\sigma_{B0}$  &  70.054\,125\,1(2) \\
$\tilde\sigma_{B1}$  & -55.342\,119\,09(14) \\
$\tilde\sigma_{B2}$  &   4.188\,033\,454(7) \\
$\tilde\sigma_{B3}$  &   0.011\,67(3)\\[1ex]
$\tilde\sigma^{(5)} $      &   4.654\,69(15) \\
\hline\hline
\end{tabular}
\end{center}
\end{table}

The numerical results for all the  contributions to the nuclear magnetic shielding for $^1$H and for the particularly important cases of $^3$He$^+$ and $^3$He are shown in Table \ref{tab:shieldingstot}.  The overall uncertainty of the obtained shielding constants is well below $10^{-10}$
and comes exclusively from the unknown higher-order terms in $\alpha$ and $m/m_N$. While
the finite nuclear size effects have be  omitted here as they are much smaller than the overall uncertainty;
nevertheless, they may become significant for heavier elements. 
\newcommand{\mct}[1]{\multicolumn{2}{l}{#1}}
\begin{table}
\caption{Contributions to the shielding constant $\sigma \cdot 10^6$ for $^1$H, $^3$He$^+$, and $^3$He (from Ref. \cite{wehrli:2021}).
Quantities that are preceded by ``$\pm$'' represent the uncertainties.
$\sigma^{(2,2)}$(He) is estimated to be between  $\sigma^{(2,2)}$(He$^+$) and $2\,\sigma^{(2,2)}$(He$^+$).
The relative uncertainty of the finite nuclear mass correction $\sigma^{(4,1)}$ is estimated as $2\,m/m_N$ of $\sigma^{(4)}$.
$\sigma^{(6)}$ is partially known, see Eq.\,\eqref{eq:diracexpansion}, but we expect cancellation with the radiative correction, 
so we estimate the uncertainty originating from this contribution as $(Z\, \alpha)^2\,\sigma^{(4)}$.}
\label{tab:shieldingstot}
\begin{center}
\begin{tabular}{ld{4.9}d{4.9}d{4.11}}
    \hline \hline & \\
 & \centt{$^1$H} &  \centt{$^3$He$^+$} & \centt{$^3$He} \\[1ex]
\hline
$\sigma^{(2)}$ &17.750\,451\,5 & 35.500\,903\,0 &  59.936\,770\,5  \\
$\sigma^{(4)}$ &0.002\,546\,9 &  0.020\,375\,1 &   0.052\,663\,1  \\
$\sigma^{(5)}$ &0.000\,018\,4 &  0.000\,082\,0 &   0.000\,096\,3\\
$\sigma^{(2,1)}$ & -0.017\,603\,7& -0.013\,933\,4 &  -0.022\,511\,5  \\
$\sigma^{(2,2)}$ &  0.000\,022\,7& 0.000\,007\,1& 0.000\,010\,7(36)   \\
$\sigma^{(4,1)}$ & \pm 0.000\,002\,8 & \pm 0.000\,007\,4& \pm 0.000\,019\,2   \\
$\sigma^{(6)}$ & \pm 0.000\,000\,1 & \pm 0.000\,004\,3  & \pm  0.000\,011\,2  \\[2ex]
$\sigma\cdot10^6$& 17.735\,436(3) & 35.507\,434(9) &   59.967\,029(23)\\
\hline\hline& 
\end{tabular}
\end{center}
\end{table}

The most important, however, is the fact that the convergence of the expansions in the fine structure constant $\alpha$ is very rapid for low-$Z$ systems, 
which justifies our approach based on NRQED theory. This is the only approach that allows for the rigorous estimation of uncertainties
in the calculation of the nuclear magnetic shielding (as well as of binding energies), in contrast to methods which are based on the Dirac-Coulomb-Breit Hamiltonian. This NRQED method can be applied to other elements, which may lead to the improved determination of magnetic moments of other nuclei.
For example, the measurement of the electron magnetic moment to the shielded nuclear one in $^9$ Be$^+$ is accurate to $10^{-9}$ \cite{WINELAND1983},
allowing for the determination of the $^9$Be nuclear magnetic moment with the similar $10^{-9}$ accuracy, which is much higher than presently known \cite{ANTUSEK2013, PACHUCKI2010}. 

\section*{Acknowledgments}
D.W. thanks F. Merkt for his unconditional support to work on this project. This research was supported by National Science Center (Poland) Grant No. 2017/27/B/ST2/02459.

\appendix
\begin{widetext}
\section{Bethe log type contribution} 
We will use atomic units throughout the Appendix for simplicity of formulas. 
Equation\,\eqref{eq:hee5la} is only a formal expression for $E_{LA}^{(5)}$, and it needs to be expanded in magnetic fields. 
For this we have to return to the integral representation (compare Eq.\,\eqref{eq:e5linterm} and the following derivation),
\begin{align}
  E^{(5)}_{LA} =&\ \frac{2\,\alpha}{3\,\pi}\int_{0}^{\epsilon}dk\,k^3\left\langle(\vec{r}_1+\vec{r}_2)\,\frac{1}{E-H-k}\,(\vec{r}_1+\vec{r}_2)\right\rangle_{\!B},
\end{align}
where in the above integral it is assumed that in the limit of large $\epsilon$, the linear and $\ln(2\,\epsilon/(m\,\alpha^2))$ terms are dropped. 

The first part $\tilde \sigma_{A1}$, due to the perturbation $1/3\,\vec{\mu}\cdot\vec{B}\, U$  from  Eq.\,\eqref{98}, is given by
\begin{align}
  \tilde{\sigma}^{(5)}_{A1} =&\ -\frac{2}{9\,\pi}\,\delta_{U}\left\langle(\vec r_1+\vec r_2)\,(H_0-E_0)^3\ln [2\,(H_0-E_0)]\,(\vec r_1+\vec r_2)\right\rangle
  \nonumber  \\ =&\
  -\frac{2}{9\,\pi}\,\delta_{U}\left\langle(\vec p_1+\vec p_2)\,(H_0-E_0)\ln[2\,(H_0-E_0)]\,(\vec p_1+\vec p_2)\right\rangle \nonumber \\
  =&\   \frac{2}{9\,\pi}\,{\cal N}_{A1},
  \label{eq:sigma5a1outline}
\end{align}
where
\begin{align}
{\cal N}_{A1} =&\ \int_{0}^{\epsilon}dk\,f^{A1},\\
f^{A1} =&\ k\,\bigg[2\left\langle U\, \frac{1}{(E_0-H_0)'}\,
 (\vec{p}_1+\vec{p}_2)\,\frac{1}{E_0-H_0-k}\,(\vec{p}_1+\vec{p}_2)\right\rangle 
 \nonumber\\ &\ 
 +\left\langle(\vec{p}_1+\vec{p}_2)\,\frac{1}{E_0-H_0-k}\,(U-\langle U\rangle)\,\frac{1}{E_0-H_0-k}\,(\vec{p}_1+\vec{p}_2)\right\rangle\bigg].
\end{align}
Changing the integration variable to $ t = 1/\sqrt{1 + 2\,k}$, ${\cal N}_{A1}$ can be rewritten in the form \cite{pachucki04}
\begin{align}
{\cal N}_{A1} =&\ \int_{0}^{1}dt\, \frac{f^{A1}(t) - f^{A1}_0 - f^{A1}_2\,t^2}{t^3},
\end{align}
where $f^{A1}_0$ and $f^{A1}_2$ are first terms in the small $t$ expansion of $f^{A1}$, namely
\begin{align}
f^{A1}_0 =&\ -2\,\left\langle U\, \frac{1}{(E_0-H_0)'}\, (\vec{p}_1+\vec{p}_2)^2 \right\rangle,  \\
f^{A1}_2 =&\  -2\,{\cal D}_{A1}, 
\end{align}
where ${\cal D}_{A1}$ is defined in Eq.\,\eqref{101}.
It is more convenient, from the numerical point of view, to consider the ratio
\begin{align}
 \ln k_{A1} = \frac{{\cal N}_{A1}}{{\cal D}_{A1}}\,, 
  \end{align}
and the $\tilde{\sigma}^{(5)}_{A1}$ low-energy contribution becomes
\begin{align}  
  \tilde \sigma^{(5)}_{A1} = \frac{2}{9\,\pi}\,{\cal D}_{A1}\,\ln k_{A1}.
\end{align}
$\ln k_{A1}$ and $\ln k_{A2}$ calculated below, similarly to the standard Bethe logarithm $\ln k_0$, exhibit the striking property
that they only weakly depend on the number of electrons (see also Ref. \cite{ferenc2020}). Table  IV presents their accurate values, which differ only slightly from the corresponding hydrogenic limits,
which are also shown in this Table.

The second part is due to perturbation from the two other terms in Eq.\,\eqref{98},
\begin{align}
  \tilde{\sigma}^{(5)}_{A2} =&\ -\frac{1}{9\,\pi}\,\delta_{L^i}\delta_{U^i}
  \left\langle(\vec r_1+\vec r_2)\,(H_0-E_0)^3\ln[2\,(H_0-E_0)]\,(\vec r_1+\vec r_2)\right\rangle\\
  =&\ 
  \frac{1}{9\,\pi}\int_{0}^{\epsilon}dk\,k^3\bigg[
    2\left\langle U^i\,\frac{1}{E_0-H_0}\,(r_1^j+r^j_2)\,\frac{1}{E_0-H_0-k}\,L^i\,\frac{1}{E_0-H_0-k}\,(r^j_1+r^j_2)\right\rangle \nonumber \\
    &\ +2\left\langle(r^j_1+r^j_2)\,\frac{1}{E_0-H_0-k}\,U^i\,\frac{1}{E_0-H_0-k}\, L^i \frac{1}{E_0-H_0-k}\,(r^j_1+r^j_2)\right\rangle\bigg]\\
  =&\ 
  \frac{1}{9\,\pi}\int_{0}^{\epsilon}dk\,k^3\,i\,\epsilon^{ijk}\bigg[
    2\left\langle U^i\,\frac{1}{E_0-H_0}\,(r_1^j+r^j_2)\,\frac{1}{(E_0-H_0-k)^2}(r^k_1+r^k_2)\right\rangle \nonumber \\
    &\ +2\left\langle(r^j_1+r^j_2)\,\frac{1}{E_0-H_0-k}\,U^i\,\frac{1}{(E_0-H_0-k)^2}\,(r^k_1+r^k_2)\right\rangle\bigg]\\ 
    =&\ 
  \frac{1}{9\,\pi}\,i\,\epsilon^{ijk}\int_{0}^{\epsilon}dk\,k^3\frac{d}{dk}\bigg[
    2\left\langle U^i\,\frac{1}{E_0-H_0}\,(r_1^j+r^j_2)\,\frac{1}{E_0-H_0-k}\,(r^k_1+r^k_2)\right\rangle \nonumber \\
    &\ +\left\langle(r^j_1+r^j_2)\,\frac{1}{E_0-H_0-k}\,U^i\,\frac{1}{E_0-H_0-k}\,(r^k_1+r^k_2)\right\rangle\bigg]\\ =&\
  \frac{1}{9\,\pi}\,( {\cal C}_{A2} - 3\,{\cal N}_{A2}),
\end{align}
where
\begin{align}
 {\cal C}_{A2}  =&\ \lim_{\epsilon\rightarrow\infty} i\,\epsilon^{ijk}\,\epsilon^3\, \bigg[
    2\left\langle U^i\,\frac{1}{E_0-H_0}\,(r_1^j+r^j_2)\,\frac{1}{E_0-H_0-\epsilon}\,(r^k_1+r^k_2)\right\rangle \nonumber \\
    &\ +\left\langle(r^j_1+r^j_2)\,\frac{1}{E_0-H_0-\epsilon}\,U^i\,\frac{1}{E_0-H_0-\epsilon}\,(r^k_1+r^k_2)\right\rangle\bigg] 
    \nonumber \\  =&\  -{\cal D}_{A2},
 \end{align}
 and
 \begin{align}
{\cal N}_{A2} =&\ \lim_{\epsilon\rightarrow\infty} \,\int_{0}^{\epsilon}dk\,f^{A2},\\
f_{A2} =&\  i\,\epsilon^{ijk}\,k^2\bigg[
    2\left\langle U^i\,\frac{1}{E_0-H_0}\,(r_1^j+r^j_2)\,\frac{1}{E_0-H_0-k}\,(r^k_1+r^k_2)\right\rangle \nonumber \\
    &\ +\left\langle(r^j_1+r^j_2)\,\frac{1}{E_0-H_0-k}\,U^i\,\frac{1}{E_0-H_0-k}\,(r^k_1+r^k_2)\right\rangle\bigg],
\end{align}
where $ {\cal D}_{A2} $ is defined in Eq.\,\eqref{104}.
Changing the integration variable to $ t = 1/\sqrt{1 + 2\,k}$, ${\cal N}_{A2}$ can be rewritten to the form 
\begin{align}
{\cal N}_{A2} =&\ \int_{0}^{1}dt\, \frac{f^{A2}(t) - f^{A2}_0 - f^{A2}_2\,t^2}{t^3},
\end{align}
where $f^{A2}_0$ and $f^{A2}_2$ are the first terms in the small $t$ expansion of $f^{A2}$, namely
\begin{align}
f^{A2}_0 =&\ \frac{2}{m}\left\langle U^i\,\frac{1}{E_0-H_0}\,(\vec r_1\times \vec p_2 + \vec r_2\times \vec p_1)^i \right\rangle + 2\left\langle\bigg(\frac{\vec r_1}{r_1^3} + \frac{\vec r_2}{r_2^3} \bigg)\cdot (\vec r_1 + \vec r_2 )\right\rangle, \\
f^{A2}_2 =&\  -2\,{\cal D}_{A2}.
\end{align}
It is more convenient, from the numerical point of view, to consider the ratio
\begin{align}
\ln k_{A2} \equiv&\ \frac{{\cal N}_{A2}}{{\cal D}_{A2}}\,, 
  \end{align}
and the $\tilde{\sigma}^{(5)}_{A2}$ low-energy contribution becomes  
\begin{align}  
  \tilde \sigma^{(5)}_{A2} =&\ -\frac{1}{9\,\pi}\, {\cal D}_{A2}\,(1+3\,\ln k_{A2}).
\end{align}
\end{widetext}

%

\end{document}